\documentclass[a4paper]{aa}

\usepackage{graphics,epsfig,txfonts}
\usepackage[section]{placeins}
\usepackage{natbib}
\bibpunct{(}{)}{;}{a}{}{,}

\newcommand{\ltsima}{$\buildrel < \over \sim$}
\newcommand{\lsim}{\lower.5ex\hbox{\ltsima}}
\newcommand{\gtsima}{$\buildrel > \over \sim$}
\newcommand{\gsim}{\lower.5ex\hbox{\gtsima}}
\usepackage{color}

\begin{document}
 
\title{7.1\,keV sterile neutrino constraints from X-ray observations of 33\,clusters of galaxies with \emph{Chandra} ACIS}

\author{      F.~Hofmann\inst{1}
     \and     J.S.~Sanders\inst{1}
     \and     K.~Nandra\inst{1}
     \and     N.~Clerc\inst{1}
     \and     M.~Gaspari\inst{2,3}
       }

\titlerunning{Sterile neutrino line limits}
\authorrunning{Hofmann et al.}

\institute{Max-Planck-Institut f\"ur extraterrestrische Physik, Giessenbachstra{\ss}e, 85748 Garching, Germany
           \and Department of Astrophysical Sciences, Princeton University, Princeton, NJ 08544, USA
           \and Einstein and Spitzer Fellow
           }


\abstract{Recently an unidentified emission line at 3.55\,keV has been detected in X-ray spectra of clusters of galaxies. The line has been discussed as a possible decay signature of 7.1\,keV sterile neutrinos, which have been proposed as a dark matter candidate.}
	 {We aim at putting constraints on the proposed line emission in a large sample of \emph{Chandra}-observed clusters and obtain limits on the mixing-angle in a 7.1\,keV sterile neutrino dark matter scenario.}
 	 {For a sample of 33 high-mass clusters of galaxies we merge all observations from the \emph{Chandra} data archive. Each cluster has more than 100\,ks of combined exposure. The resulting high signal-to-noise spectra are used to constrain the flux of an unidentified line emission at 3.55\,keV in the individual spectra and a merged spectrum of all clusters.}
 	 {We obtained very detailed spectra around the 3.55\,keV range and limits on an unidentified emission line. Assuming all dark matter were made of 7.1\,keV sterile neutrinos the upper limits on the mixing angle are $\rm{sin^2(2\Theta)}$ $\rm{<10.1\times10^{-11}}$ from ACIS-I, and $\rm{<40.3\times10^{-11}}$ from ACIS-S data at 99.7 per cent confidence level.}
	 {We do not find evidence for an unidentified emission line at 3.55\,keV. The sample extends the list of objects searched for an emission line at 3.55\,keV and will help to identify the best targets for future studies of the potential dark matter decay line with upcoming X-ray observatories like \emph{Hitomi (Astro-H)}, \emph{eROSITA}, and \emph{Athena}.}


\maketitle

\section{Introduction}
\label{sec:introduction}

\citet{2014ApJ...789...13B} and \citet{2014PhRvL.113y1301B} recently found indications for a weak unidentified emission line ($\rm{E \sim 3.55~keV}$) in X-ray CCD spectra of the Andromeda galaxy and in deep cluster observations using \emph{Chandra} and \emph{\emph{XMM-Newton}} data. There is an ongoing discussion on the existence and possible nature of the line with studies using other instruments like the \emph{Suzaku} observatory \citep[][]{2015MNRAS.451.2447U} and looking at other objects like individual galaxies \citep[e.g.][]{2015MNRAS.452.3905A}. The line has been proposed as a candidate for a dark matter (DM) decay line and could be explained by decay of sterile neutrinos with a mass of $\rm{m_s=7.1~keV}$. In this model they decay into an X-ray photon with $\rm{E_{\gamma}=m_s/2}$ and an active neutrino $\rm{\nu}$. Sterile neutrinos with masses in the keV range have long been discussed as a possible component of DM \citep[e.g.][]{1994PhRvL..72...17D, 2001ApJ...562..593A, 2009ARNPS..59..191B}, but up until recently only upper limits could be derived \citep[e.g. from observations of the Andromeda galaxy or the Bullet Cluster by][]{2008MNRAS.387.1361B, 2008ApJ...673..752B}.

In the case of sterile neutrino decay the measured additional flux at $\rm{\sim3.55~keV}$ would be related to two defining properties of the particles: The particle mass $\rm{m_s}$ and the mixing-angle $\rm{sin^2(2\Theta)}$, which describes interaction of the sterile neutrinos with its active neutrino counter-parts and thus the likelihood of decay in the $\rm{\gamma/\nu}$ channel. They are related through \citep[as used by][]{2014ApJ...789...13B}

\begin{equation}
 \rm{sin^2(2\Theta) = \frac{F_{DM}~10^{14}~M_{\odot}}{\frac{12.76}{cm^2~s}~M^{FOV}_{DM}~(1+z)}}~\left(\frac{D_L}{100~Mpc}\right)^2\left(\frac{1~keV}{m_s}\right)^4
\label{eqn:sin}
\end{equation}

where $\rm{F_{DM}}$ is the observed flux of the DM decay line, $\rm{M^{FOV}_{DM}}$ is the expected DM mass within the field of view (FOV) of the observation, $\rm{D_L}$ is the luminosity distance of the cluster, and z its redshift.

The paper is structured as follows: Sect. 2 describes the sample of clusters and the data reduction, Sect. 3 describes the fitting of individual cluster spectra, in Sect. 4 the procedure for fitting merged spectra is introduced, Sect. 5 explains how DM masses in the FOV were estimated, in Sect. 6 the upper limits for an additional 3.55\,keV line emission are presented, in Sect. 7 the results are discussed comparing to other studies, and in Sect. 8 our findings are summarized.

For all our analysis we used a standard $\rm{\Lambda CDM}$ cosmology with $\rm{H_0=71~km~s^{-1}~Mpc^{-1}}$, $\rm{\Omega_M=0.27}$ and $\rm{\Omega_{\Lambda}=0.73}$ and relative solar abundances as given by \citet{1989GeCoA..53..197A}.

\section{Observations and data reduction}
\label{sec:observations}

We studied X-ray spectra of a sample of 33 clusters of galaxies. For a detailed description of the sample selection and data reduction see \citet{2016A&A...585A.130H}. They were selected as the most X-ray luminous clusters with more than 100\,ks raw exposure with the \emph{Chandra} Advanced CCD Imaging Spectrometer \citep[ACIS, energy range about 0.1 to 10\,keV,][]{2003SPIE.4851...28G}. 
The combined exposures of the sample are $\rm{\sim5.7~Ms}$ for ACIS-I and $\rm{\sim2.3~Ms}$ for the ACIS-S detectors.
The redshift range of the sample is $\rm{0.025 < z < 0.45}$ with a mass range of $\rm{1\times10^{14}~M_{\odot}}$ to $\rm{2\times10^{15}~M_{\odot}}$. 

The instruments spectral resolution is $\rm{\sim110-150~eV}$ which is broader than the energy difference to some neighbouring emission lines in the 3.55\,keV region so their influence has to be carefully modelled (see discussion in Sect. \ref{sec:discussion}).
We obtained the observational data from the \emph{Chandra} data archive and reprocessed them using the standard data processing. We added all available observations for each cluster to obtain the deepest images of the systems. On the reduced images a spatial-spectral extraction was performed by using the contour binning technique \texttt{contbin} \citep[see][]{2006MNRAS.371..829S}, which divides the cluster emission into smaller regions of equal signal-to-noise (here S/N = 50 or 25 depending on data quality). Based on the obtained maps of the cluster we added all spectra above a fixed surface brightness where the cluster emission is homogeneously covered by all observations and obtained deep X-ray spectra for each cluster (separately for ACIS-I/-S). For a detailed description of the data reduction see \citet{2016A&A...585A.130H}. 

\section{Fitting individual spectra}
\label{sec:ind_spec}

\begin{figure}
 \centering
 \includegraphics[width=0.48\textwidth,angle=0,trim=0.7cm 0cm 0cm 0cm,clip=true]{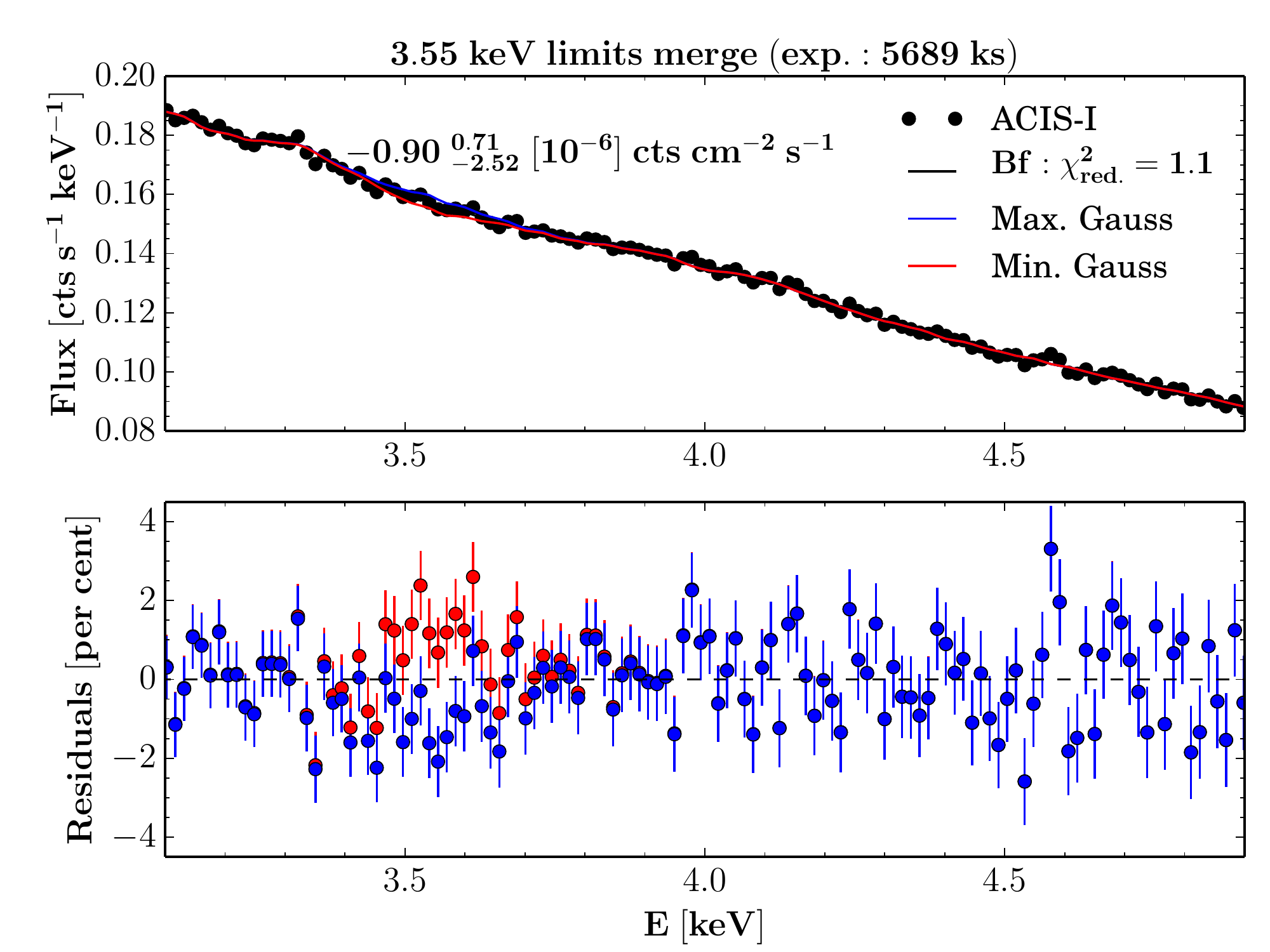}
 \includegraphics[width=0.48\textwidth,angle=0,trim=0.7cm 0cm 0cm 0cm,clip=true]{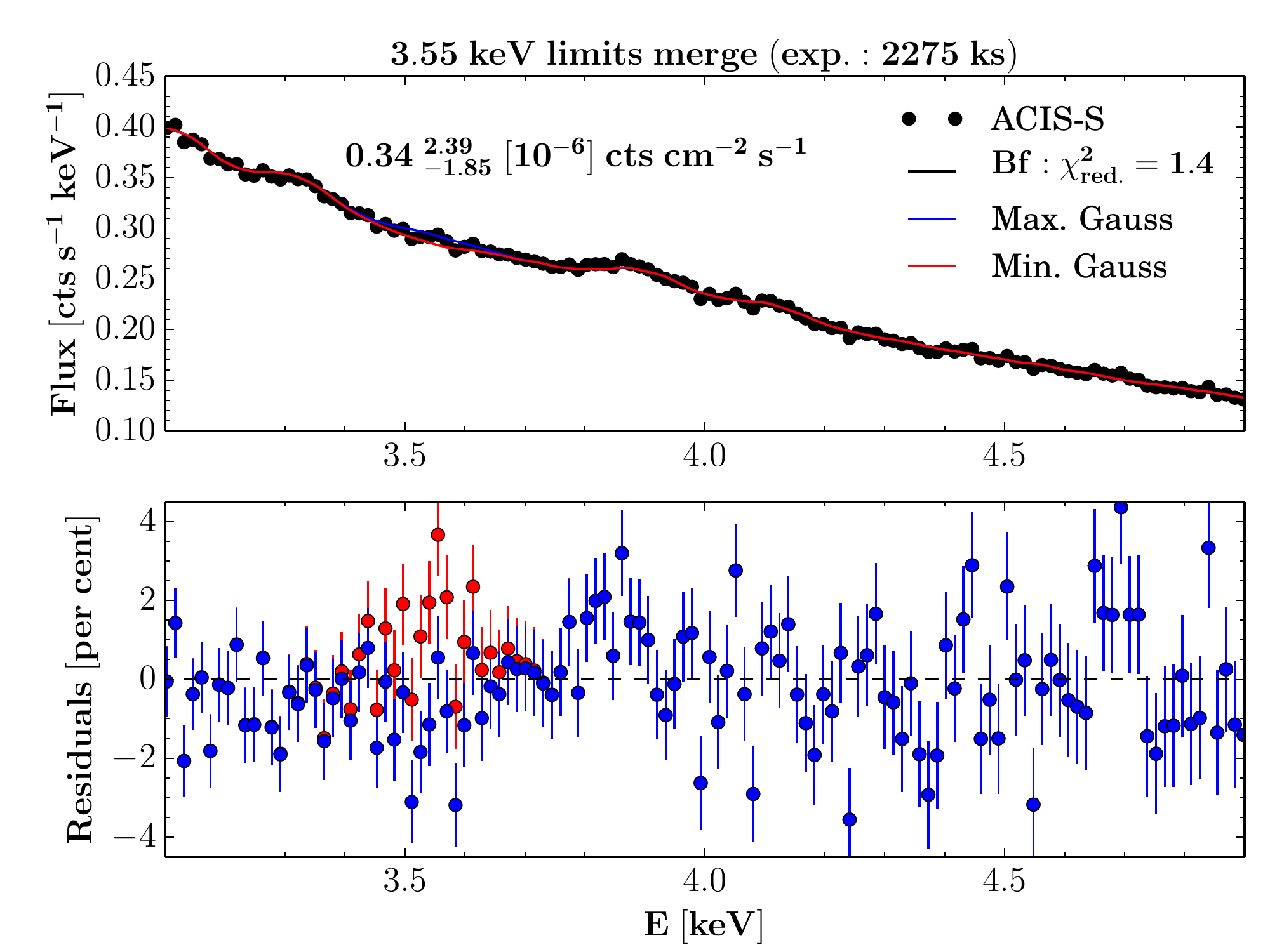}
 \caption{Merged X-ray spectra (ACIS-I top and ACIS-S bottom) of the cluster sample with residuals of different fitted models. Fitted \texttt{XSPEC} models: \texttt{apec+apec+zgauss} with best-fit (Bf), upper and lower confidence values (99.7 per cent) of the Gaussian flux in counts $\rm{cm^{-2}~s^{-1}}$. The annotations show the best-fit value and the confidence interval obtained using MCMC. Residuals are shown for the fit with upper (blue) and lower (red) confidence limit of the Gaussian flux. For the effective-area curve see Fig. \ref{fig:fakelinemerge}.}
 \label{fig:linemerge}
\end{figure}

For every cluster we added the source and background spectra of every spatial bin from the maps (see Sect. \ref{sec:observations}, only inner regions with high surface brightness and homogeneous coverage for both ACIS-I/-S). The background was renormalised to match the count rate in the 10.0-12.5\,keV energy range. The response files were averaged and weighted by the number of counts in the spectrum (both auxiliary response files, ARF, and redistribution matrix files, RMF). For analysing the spectra we used \texttt{XSPEC} version 12.9.0 \citep[][]{1996ASPC...99..409A} and ATOMDB version 2.0.2 \citep[][]{2012ApJ...756..128F}.

To estimate the upper limit of the flux allowed for an additional emission line, we searched for the best fitting \texttt{apec} model (with two temperature components) for collisionally-ionized plasma with absorption ($\rm{n_H}$) and an additional zero-width Gaussian line redshifted corresponding to the cluster distance (see Fig. \ref{fig:3.55all} for spectra of individual clusters). 
The normalisation of the Gaussian was allowed to be negative to avoid bias.
All spectra were grouped to contain a minimum of 22 raw counts in each bin (using \texttt{grppha}) and we used the range from 2-5\,keV for fitting the spectral model to the data (using $\rm{\chi^2}$ statistics). Free parameters of the fit were the normalisation of the spectral components, the temperatures, and the relative abundances of the \texttt{apec} models. 
The fit was done using a fixed foreground column density ($\rm{n_H~[cm^{-2}]}$), which was determined from the Leiden/Argentine/Bonn (LAB) survey of Galactic HI \citep[based on][]{2005A&A...440..775K}. Where we found the absorption to vary significantly across the FOV of an observation $\rm{n_H}$ was set as additional free parameter of the fit \citep[only for pks0745, see][]{2016A&A...585A.130H}. Additional absorption of about 5 per cent can be caused by molecular hydrogen in the foreground \citep[$\rm{H_2}$ X-ray absorption, see][]{2013MNRAS.431..394W}. We did not include the additional absorption by $\rm{H_2}$ because the influence on the spectral form in the 3\,keV range is negligible.
If the two temperatures were separated by less than 30 per cent or if the normalisation of one component was less than one per cent of the second we discarded the second component and fit only a one-temperature \texttt{apec} model (see Tab. \ref{tab:xspecpars}).

Once the best fit was identified, we calculated the confidence intervals (99.7 per cent) for the additional flux added by the Gaussian using a Monte Carlo Markov Chain (MCMC) method with length of 10000 and burn-in length of 1000. As a robustness test we used a Bayesian fitting code \citep[BXA;][]{2014A&A...564A.125B} with flat priors for the same parameter ranges as our standard fitting method and obtained consistent results. A non-negative prior for the additional Gaussian flux did not improve the constraints.

\section{Fitting merged spectra}
\label{sec:merge_spec}

\begin{figure*}
 \centering
 \includegraphics[width=\textwidth,angle=0,trim=0cm 3.8cm 0cm 0cm,clip=true]{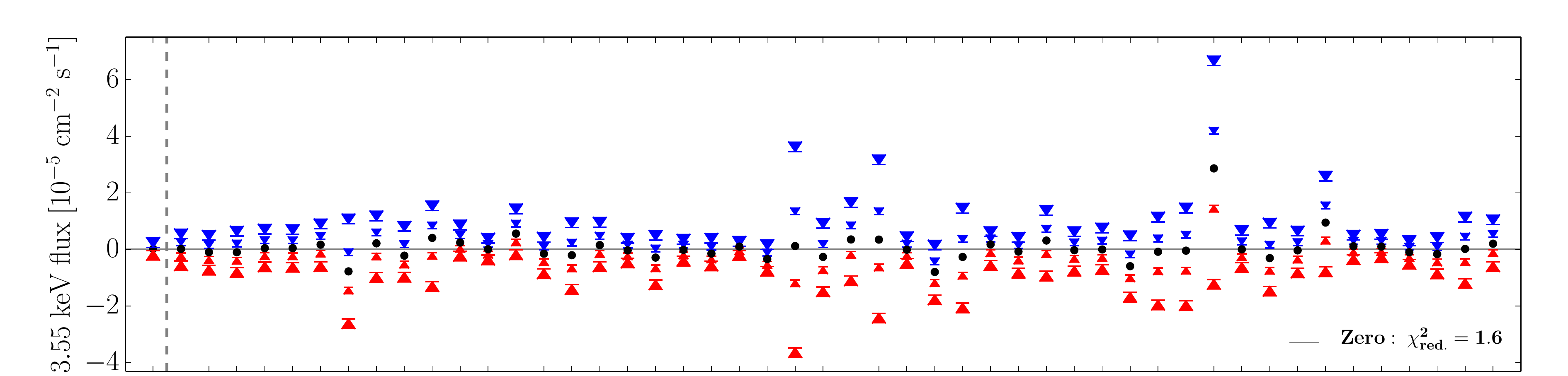}
 \includegraphics[width=\textwidth,angle=0,trim=0cm 0cm 0cm 0.6cm,clip=true]{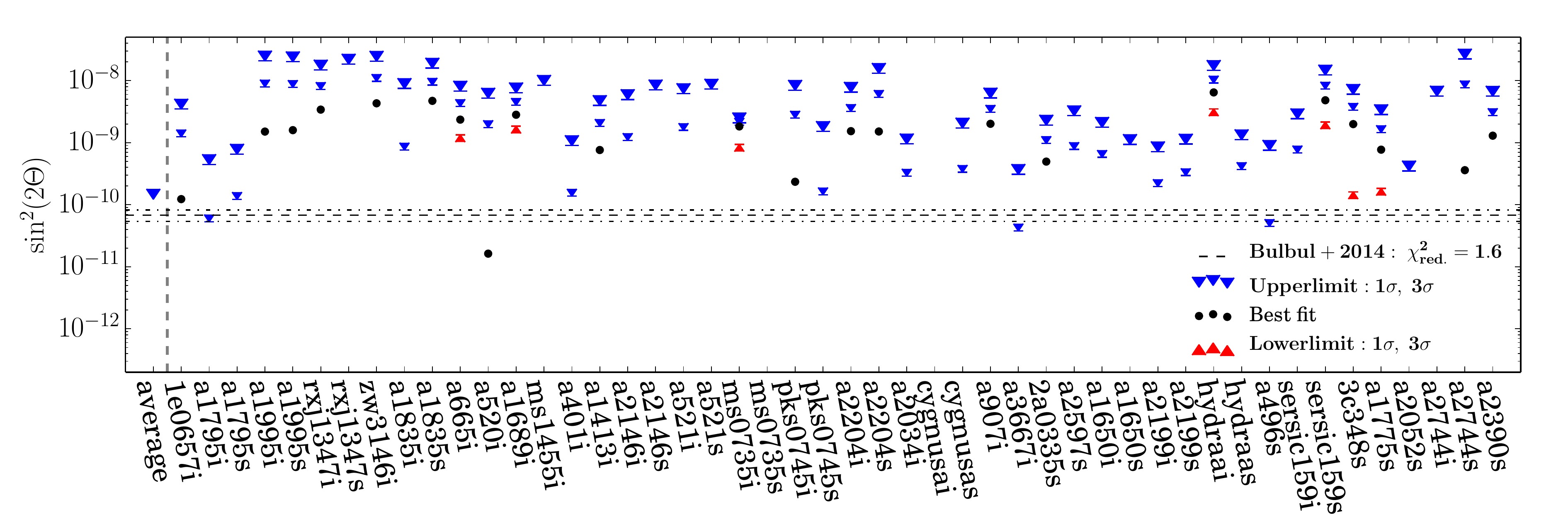}
 \caption{Top: limits on an additional Gaussian flux component at 3.55\,keV for all clusters in the sample. Bottom: limits on the mixing-angle in case of a 7.1\,keV sterile neutrino. The limits are separately calculated for ACIS-S and ACIS-I, which is indicated by -s or -i following the cluster identification on the x-axis. Most lower limits and some best fit values lie outside the plotted range. The dashed line shows the \citet{2014ApJ...789...13B} detection in their full \emph{\emph{XMM-Newton}} MOS sample. Limits show the $\rm{1\sigma}$ (small symbols) and $\rm{3\sigma}$ (big symbols) confidence range.}
 \label{fig:lineall}
\end{figure*}

To obtain deeper limits on the line we added all cluster spectra in our sample. We de-redshifted all spectra by re-binning the counts to the channel-grid of the new spectra after randomising each count position within the channel width.
The ARFs were de-redshifted by calculating the new energy values of the spectral channels and then mapping back to the original channel-grid by interpolating the shifted response. For the RMF we treated the integer value of the response matrix in each pixel (2D detector response matrix for channel to energy mapping) as a number of counts and applied the same de-redshifting procedure with randomisation as to the spectra. To approximate the new response files better we repeated the process 100 times and calculated the average. Residual noise from the re-mapping (caused by the discrete energy grid) was reduced by smoothing the responses with a Gaussian filter ($\rm{\sigma=1~pixel}$). The width of the energy bins is $\rm{\sim 0.01~keV}$ for both detectors (see response in Fig. \ref{fig:fakelinemerge}).
The de-redshifting was based on optical redshifts known from previous catalogues (see Tab.\ref{tab:clusterpars}).
We added all de-redshifted, unabsorbed spectra and averaged the response files with weights according to the fraction of the total counts of the merged spectrum (in the 2-5\,keV range). 
In the fitted range 2-5\,keV, the merged spectra contain $\rm{3.2\times10^6}$ counts (ACIS-I) and $\rm{2.7\times10^6}$ counts (ACIS-S).
The average exposure times of each observation were added to obtain an average count rate in the final spectra. The background spectrum was renormalised to match the 10.0-12.5\,keV energy range count rate.
We used the \texttt{XSPEC} models \texttt{apec+apec+zgauss} without absorption and z=0 to fit the spectra and estimate the additional flux allowed for a line at 3.55\,keV. 
The fit of the merged spectra was performed between 3-5\,keV because there is no strong variation in the detector response in that region and thus possible systematics in the calculated average response are minimal.
Fig. \ref{fig:linemerge} shows some residual structures due to the modelling of a complex stack of cluster spectra with a two-temperature \texttt{apec} model. 

We tested fitting the emission features in the spectrum with separate Gaussian lines on top of an \texttt{apec} model without lines \citep[compare e.g.][]{2014ApJ...789...13B}. 
This procedure provided a goodness of fit comparable to the fitting process used in this study. We concluded that the modelling with two temperature collisionally ionised plasma emission (\texttt{XSPEC} \texttt{apec} model) is appropriate for the spectra in this study. 

There are some remaining residuals between the model and data in the merged spectra. A large fraction of these residuals are likely caused by the systematic effects of averaging and de-redshifting the responses and spectra (see simulated spectra in Fig. \ref{fig:fakelinemerge}). As these features are narrower than the response, they cannot and should not be fitted by Gaussian lines. Modelling the emission lines separately \citep[e.g. as][]{2014ApJ...789...13B} does not significantly improve the fit, except for the two strongest features at 3.85 and 4.7\,keV (perhaps due to incomplete modelling of the temperature structure, as discussed in Appendix \ref{sec:residuals}). If these two features are included in our model, there is no significant change to our limits. We conclude that the two temperature plasma model is appropriate for this study. 
There are no strong emission lines expected over the range in temperature in the region of the 3.55\,keV line, which could explain the strength of previous detections.

Fig. \ref{fig:linemerge} shows the additional residuals at $\rm{\sim3.85~keV}$ and $\rm{\sim4.7~keV}$. The shape of spectral features in the CCD spectra can change with temperature because they are blends of several neighbouring emission lines and the ratio between those is temperature dependent.

We used simulations to verify the line-detection efficiency of our stacking and modelling method (see Appendix \ref{sec:simspec}) and show that small additional residuals from modelling the many averaged temperatures of the ICM with a two temperature model does not affect the limits derived for the 3.55\,keV line (see Appendix \ref{sec:residuals}).

\section{Estimating cluster masses}
\label{sec:mass}

To constrain the expected strength of the DM emission line in the cluster spectra we need to estimate the DM mass in the FOV. With the measured average temperatures of the clusters we used temperature-mass scaling relations from \citet{2009ApJ...692.1033V} to estimate the overall cluster mass $\rm{M_{500}}$ and the related overdensity radius $\rm{r_{500}}$ \citep[see][for details]{2016A&A...585A.130H}. 

We assumed the cluster density follows an NFW profile \citep[][]{1997ApJ...490..493N} and used a fixed concentration (C=4), as the expected average value of the sample \citep[compare e.g.][]{2006ApJ...640..691V}. 
By integrating $\rm{\rho(r)}$ along the line of sight we obtained the DM mass surface density profiles \citep[see e.g.][]{2000ApJ...534...34W,2015MNRAS.451.2447U}. 
We estimated the DM mass in the FOV of an observation by adding up the surface densities within the footprint of the extraction region for the spectra. The footprints of each observation and the maximum covered radius have been published by \citet{2016A&A...585A.130H}.

For \object{A\,1795} we obtained an $\rm{M^{FOV}_{DM}}$ of $\rm{1.6\times10^{14}~M_{\odot}}$ \citep[$\rm{2.75\times10^{14}~M_{\odot}}$ for $\rm{M_{2500}}$ given by][]{2006ApJ...640..691V}. We estimate the systematic uncertainties of $\rm{M^{FOV}_{DM}}$ to be 35 per cent on average due to scatter in the scaling relations for the total mass and scatter around the average concentration. The scatter will average out when adding together properties of the whole sample which means that we expect much smaller systematic uncertainties in the average $\rm{M^{FOV}_{DM}}$ used for limits on the merged cluster spectra (see Tab. \ref{tab:clusterpars}). 

\section{Upper limits on 3.55\,keV line}
\label{sec:lim}

We investigated the possible existence of an unidentified X-ray emission line at $\rm{\sim3.55~keV}$ in the cluster spectra. We put limits on the existence of such a line in all the clusters individually and in a merged spectrum of the whole sample. A possible interpretation for this line has been the emission from decaying sterile neutrinos with a mass of $\rm{\sim7.1~keV}$ which could be a candidate for DM.

\subsection{Limits in individual spectra}
\label{sec:ind_lim}

Fig. \ref{fig:lineall} shows the upper limits on the flux derived from observations of A\,1795 (deepest observation in the sample), where we obtained a $\rm{3\sigma}$ upper limit of $\rm{3.22\times10^{-6}~cm^{-2}~s^{-1}}$ from merged ACIS-I observations ($\rm{\sim613~ks}$) and $\rm{4.73\times10^{-6}~cm^{-2}~s^{-1}}$ from ACIS-S ($\rm{\sim241~ks}$).

Using equation (\ref{eqn:sin}) this translates to one of the deepest upper limits on the mixing-angle for individual clusters $\rm{sin^2(2\Theta) \leq 4.4\times10^{-10}}$ (ACIS-I data).
The limits on the additional flux that we could derive in the individual systems depend on the exposure time and on how well the cluster emission is modelled by a two component collisionally-ionised plasma model. 
For a detailed list of constraints see Tab. \ref{tab:indi_limits}.

\subsection{Limits in merged spectra}
\label{sec:merge_lim}

The fits to the merged data provide deeper upper limits on the flux by about a factor of three for ACIS-I compared to the individual spectra (see Fig. \ref{fig:linemerge} and Tab. \ref{tab:mix}). To turn this flux into a limit on the expected mixing-angle in the 7.1\,keV sterile neutrino scenario we estimated the average mass and luminosity distance of the cluster sample. We weighted the contributions to the average with their expected contribution $\rm{\omega_{i,DM}}$ to the DM line flux with the formula used by \citet{2014ApJ...789...13B},

\begin{equation}
 \rm{\omega_{i,DM} = \frac{M_{i,DM}(1+z_i)}{4\pi~D^2_{i,L}}\times\frac{exp_i}{exp_{tot}}}
\label{eqn:weights}
\end{equation}

where i denotes the properties of the ith cluster and $\rm{exp_i/exp_{tot}}$ is the fraction of the total exposure for the ith cluster. 
This is done for ACIS-I and ACIS-S separately (see weights in Tab. \ref{tab:clusterpars}).

The most stringent limit on the mixing-angle from this analysis is $\rm{sin^2(2\Theta) \lesssim 10\times10^{-11}}$ (99.7 per cent confidence) for the merged ACIS-I spectra of the sample. All $\rm{3\sigma}$ (99.7 per cent) upper limits are compatible with the value of $\rm{sin^2(2\Theta) \sim 7\times10^{-11}}$ found for the full sample of \citet{2014ApJ...789...13B}. A simultaneous fit of the merged ACIS-I and ACIS-S spectrum did not improve the limits.

\section{Discussion}
\label{sec:discussion}
\begin{table*}[]
\caption[]{Limits from merged spectra.}
\begin{center}
\begin{tabular}{lrrrrrr}
\hline\hline\noalign{\smallskip}
  \multicolumn{1}{l}{Merged Spec.} &
  \multicolumn{1}{l}{$\rm{F_X}$\tablefootmark{a}}&
  \multicolumn{1}{l}{$\rm{sin^2(2\Theta)}$\tablefootmark{b}} &
  \multicolumn{1}{l}{$\rm{<n_H>}$\tablefootmark{c}} &
  \multicolumn{1}{l}{$\rm{<z>}$\tablefootmark{c}} &
  \multicolumn{1}{l}{$\rm{<M_{DM}^{FOV}>}$\tablefootmark{c}} &
  \multicolumn{1}{l}{$\rm{<M_{DM}^{FOV}/D^{2}>}$\tablefootmark{c}} \\[1pt]
\multicolumn{1}{l}{} &
  \multicolumn{1}{l}{$\rm{[10^{-6}~cm^{-2}~s^{-1}]}$}&
  \multicolumn{1}{l}{$\rm{[10^{-11}]}$} &
  \multicolumn{1}{l}{$\rm{[10^{22}~cm^{-2}]}$} &
  \multicolumn{1}{l}{$\rm{}$} &
  \multicolumn{1}{l}{$\rm{[10^{14}~M_\odot]}$} &
  \multicolumn{1}{l}{$\rm{[10^{10}~M_\odot~Mpc^{-2}]}$} \\
\noalign{\smallskip}\hline\noalign{\smallskip}

mergei & $\rm{-0.9^{+1.6}_{-1.6}}$ &  $\rm{-12.7^{+22.8}_{-22.9}}$ &  0.06 & 0.11 & 2.08 & 0.197 \\[2pt]
merges & $\rm{0.3^{+2.1}_{-2.2}}$ &  $\rm{5.7^{+34.6}_{-37.0}}$ &  0.10 & 0.07 & 1.26 & 0.172 \\

\noalign{\smallskip}\hline
\end{tabular}
\tablefoot{
\tablefoottext{a}{Best fit and $\rm{3\sigma}$ confidence range of a possible additional flux added by a Gaussian line at 3.55\,keV.}
\tablefoottext{b}{Mixing-angle limits ($\rm{3\sigma}$ confidence) for the 7.1\,keV sterile neutrino decay scenario (negative mixing-angle unphysical, results from allowing negative flux).}
\tablefoottext{c}{Weighted average properties of the cluster sample weighting contributions as described in equation (\ref{eqn:weights}) in Sect. \ref{sec:merge_lim}.}
}
\end{center}
\label{tab:mix}
\end{table*}
\begin{figure}
 \centering
 \includegraphics[width=0.48\textwidth,angle=0,trim=0cm 0.4cm 0cm 0.2cm,clip=true]{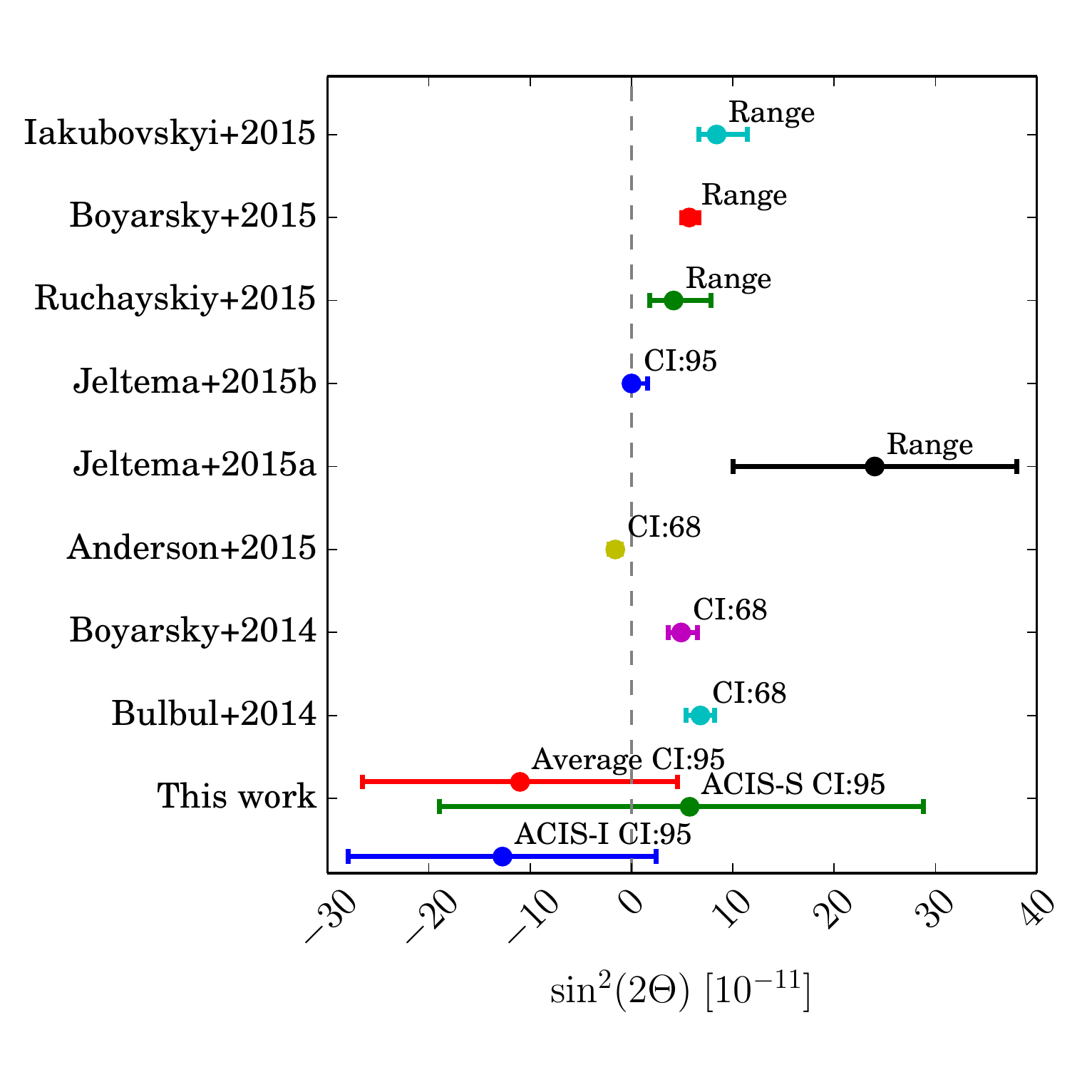}
 \caption{Comparison of this works constraints to selected literature values. The best fit line energy varies between $\rm{\sim3.4-3.6~keV}$. This work: 95\% confidence interval (CI) for \emph{Chandra} ACIS-I (bottom) and ACIS-S (middle) stacked spectra and for the error-weighted average of all individual limits (top). \citet{2014ApJ...789...13B}: 68\% CI for the full cluster sample using the \emph{XMM-Newton} MOS detectors. \citet{2014PhRvL.113y1301B}: 68\% CI (statistical only) for the M\,31 \emph{XMM-Newton} data. \citet{2015MNRAS.452.3905A}: 68\% CI for negative residual in stacked \emph{XMM-Newton} data of galaxy halos. Jeltema+2015a \citep[see][]{2015MNRAS.450.2143J}: range of values obtained for different mass models of the Galactic center with \emph{XMM-Newton} observations. Jeltema+2015b \citep[see][]{2016MNRAS.458.3592J}: 95\% CI upper limit from a deep observation of the Draco dwarf galaxy with \emph{XMM-Newton}. \citet{2015arXiv151207217R}: range of values obtained for different mass models of the Draco dwarf galaxy with the \emph{XMM-Newton} PN instrument. \citet{2015PhRvL.115p1301B}: best fit range for observations of the Galactic center with the \emph{XMM-Newton} MOS instrument. \citet{2015arXiv150805186I}: range of best fit values for a sample of galaxy clusters observed with \emph{XMM-Newton}. Negative mixing-angles are unphysical but result from equation (1) when allowing negative flux. The plot only shows selected constraints from the listed publications.}
 \label{fig:literature}
\end{figure}

We did not find evidence for an unidentified emission line at 3.55\,keV (see Fig. \ref{fig:lineall}). 
The average value of the Gaussian flux measured in the cluster sample is consistent with zero with some scatter ($\rm{\chi^2_{red.}=1.6}$), suggesting mild additional systematic uncertainties ($\rm{\sim 25}$ per cent). 
This is a conservative estimate because the high $\rm{\chi^2_{red.}}$ value is mainly driven by outliers like Hydra\,A.
The average of the calculated mixing-angle is much lower than the value of \citet{2014ApJ...789...13B}, but is consistent within the $\rm{3\sigma}$ upper limit (see Fig. \ref{fig:lineall}).

The highest $\rm{3\sigma}$ upper limit and best fit is obtained for the spectrum of a $\rm{\sim 20~ks}$ ACIS-I observation of Hydra\,A. The width of the fit line in this case is unusually broad (due to inaccuracies in the calibration files of the observation, ID 575 from 1999) and the line is not detected in a $\rm{\sim 200~ks}$ observation of the same cluster with ACIS-S (see Fig. \ref{fig:lineall} for derived limits on mixing-angle). 

A difference to the \citet{2014ApJ...789...13B} analysis is that we used only core regions of the clusters where the gas emission causes a higher background for detection of any DM line emission and we might thus be less sensitive. Using \emph{Chandra} ACIS we could not extend the analysis to larger radii due to the limited FOV. 
Another difference in the analysis is that we did not use a separately weighted response for the Gaussian DM line for modelling the merged data.

\citet{2015MNRAS.451.2447U} used deep data taken with the \emph{Suzaku} observatory and found some evidence for the predicted emission line in spectra of the Perseus cluster. However their study of the Coma, Virgo, and Ophiuchus clusters disfavour the DM nature of the line.
\citet{2015MNRAS.452.3905A} excluded the existence of the DM decay line in the merged spectra of nearby galaxy halos, where the ratio of DM to baryonic, X-ray emitting matter, is expected to be higher \citep[see also][]{2014PhRvD..90j3506M}.

The Draco dwarf spheroidal galaxy was targeted by very deep observations with the \emph{XMM-Newton} satellite. Depending on the modelling these observations are in tension with the dark matter interpretation of the line with $\rm{\sim 2\sigma}$ \citep[see][]{2016MNRAS.458.3592J}. 
In the same dataset \citet{2015arXiv151207217R} found an upper limit on the sterile neutrino lifetime which is consistent (within $\rm{1-2~\sigma}$ uncertainties) with previous detections in the scenario that the sterile neutrinos with 7.1\,keV make up all DM in the halo.
\citet{2015MNRAS.450.2143J} found evidence of a line around 3.5\,keV in \emph{XMM-Newton} data of the Galactic centre.
Other detections of the additional line include measurements of the Galactic center X-ray emission \citep[][]{2015PhRvL.115p1301B}.
The recent \emph{XMM-Newton} flux measurements for an additional line in the 3.5\,keV regime in a sample of clusters of galaxies by \citet{2015arXiv150805186I} is in weak tension with our limits (e.g. $\sim2\sigma$ for Abell\,2199).
\citet{2015ApJ...810...48R} used deep observations of the Bullet cluster with the NuSTAR satellite to constrain possible decay lines up to higher energies but due to lower sensitivity at 3.5\,keV do not exclude previous detections.

The main caveat of our mixing-angle calculations is the uncertainty of the DM masses in the FOV which are derived from scaling relations with the temperature of the hot ICM. The main caveats of the flux constraints are possible uncertainties in the calibration of the effective area and detector response and in the assumed background spectrum (derived from \emph{Chandra} blank sky observations). 
Alternative explanations for the potential line could be unexpectedly strong plasma emission lines or charge exchange processes around 3.55\,keV \citep[see discussion by][]{2014ApJ...789...13B,2015A&A...584L..11G}. Candidate plasma lines are K\,XVIII (3.476, 3.496 and 3.515\,keV), Cl\,XVII (3.510 and 3.509\,keV) as well as Ar\,XVII (3.617 and 3.618\,keV) \footnote{http://atomdb.org}. 

The merged spectra in Fig. \ref{fig:linemerge} show an additional weak ($\rm{\gtrsim1\sigma}$) residual around 4.7\,keV (mainly in ACIS-S) which is not visible in the simulations (see Fig. \ref{fig:fakelinemerge}). This could be explained by an unexpectedly high abundance (higher than assumed in \texttt{apec} metallicity model) of Titanium (Ti\,XXI, multiple lines between 4.70-4.76\,keV). The residual at 4.7\,keV disappears for 1-3 times solar abundance but in this case the line of Ti\,XXII at about 4.9\,keV is not observed at the expected strength.

Fig. \ref{fig:literature} shows a comparison of this work to previous measurements (see also discussion above). The results show conservative limits from the \emph{Chandra} ACIS-S and ACIS-I analysis and the offset shows possible systematics between instruments. The average value of all individual limits is closer to the ACIS-I measurement since the sample has higher exposure in ACIS-I and more clusters have ACIS-I observations. Most detections are consistent with our 95 per cent confidence regions. Reasons for the more conservative limits can be the smaller FOV of \emph{Chandra}, the lower combined exposure compared to some sample studies, the smaller effective area of \emph{Chandra} compared to \emph{XMM-Newton}, and a more conservative modelling approach.

\section{Conclusions}
\label{sec:concl}

This study presents a large sample of deep cluster spectra with limits on the previously detected emission at 3.55\,keV assuming the line originated from the decay of sterile neutrino DM ($\rm{m_s = 7.1~keV}$). 
We extend the number of objects previously searched for the line and provide further insight to whether the line only occurs in special observations or objects.
This is the first study using a large sample of \emph{Chandra}-observed clusters to constrain the 3.55\,keV line.
The driving cluster property behind the depth of the upper limit on the mixing-angle in the sterile neutrino scenario is the DM mass in the FOV where the spectra have been extracted divided by the luminosity distance squared. To maximize this property, homogeneous, deep coverage out to large radii of a massive nearby system is needed (higher $\rm{[10^{10}~M_\odot~Mpc^{-2}]}$ value in Tab. \ref{tab:clusterpars}). 
In this sample the best candidate for such a study with \emph{Chandra} was Abell\,1795 because of its very deep exposure (see Tab. \ref{tab:clusterpars}).

As demonstrated by \citet{2014ApJ...789...13B} a 1\,Ms observation of the Perseus cluster with the \emph{Hitomi (Astro-H)} SXS instrument \citep[][]{2014arXiv1412.1176K} will allow to distinguish between a plasma emission line of the ICM, broadened by the turbulence in the cluster ($\rm{\sim 300~km/s}$) and a DM decay line, broadened by the virial velocity ($\rm{\sim 1300~km/s}$) of the DM halo.
The large FOV of the \emph{eROSITA} observatory \citep[][]{2012arXiv1209.3114M} will allow for tight constraints on the line \citep[see also][]{2015JCAP...09..060Z}, homogeneously covering nearby X-ray bright clusters to large radii, even with lower effective area at 3.55\,keV compared to \emph{\emph{XMM-Newton}} and \emph{Chandra}. 
Only deeper observations with current or future instruments will allow to finally decide the nature of the detection as summarised in a recent white paper on keV sterile neutrinos \citep[][]{2016arXiv160204816A}.

\begin{acknowledgements}

We thank the anonymous referee for constructive comments that helped to improve the clarity of the paper. 
We thank J. Buchner, A. Del Moro, C. Pulsoni, and J. Ridl for helpful discussions.
This research has made use of data obtained from the \emph{Chandra} Data Archive, software provided by the \emph{Chandra} X-ray Center (CXC) in the application package CIAO, NASA's Astrophysics Data System, data and software provided by the High Energy Astrophysics Science Archive Research Center (HEASARC), which is a service of the Astrophysics Science Division at NASA/GSFC and the High Energy Astrophysics Division of the Smithsonian Astrophysical Observatory, the tools Veusz, the matplotlib library for Python, and TOPCAT.
M.G. is supported by the National Aeronautics and Space Administration through Einstein Postdoctoral Fellowship Award Number PF-160137 issued by the \emph{Chandra} X-ray Observatory Center, which is operated by the Smithsonian Astrophysical Observatory for and on behalf of the National Aeronautics Space Administration under contract NAS8-03060. 

\end{acknowledgements}

\bibliographystyle{aa}
\bibliography{auto,general}

\Online
\begin{appendix}

\section{Dark matter mass profiles}
\label{sec:acalc}

We assumed the cluster density follows an NFW profile \citep[][]{1997ApJ...490..493N} depending on the radius r from the centre,

\begin{equation}
 \rm{\rho(r)=\frac{\delta_c\rho_c}{r/r_s~(1+r/r_s)^2}}
\end{equation}

\begin{equation}
 \rm{\delta_c=\frac{500}{3}\times\frac{C^3}{ln(1+C)-C/(1+C)}}
\end{equation}

where concentration C and radius $\rm{r_s}$ are empirical cluster specific values. C describes how strongly peaked the density profile is in the centre and $\rm{r_s}$ is a specific radius depending on the extent of the cluster (here we define $\rm{r_s=r_{500}/C}$). $\rm{\rho_c}$ is the critical density of the universe at the cluster redshift. In the standard $\rm{\Lambda CDM}$ cosmology assumed in this study (see Sect. \ref{sec:introduction}) the critical density is given as

\begin{equation}
 \rm{\rho_c=\frac{3H^{2}(z)}{8\pi G}}
\end{equation}

\begin{equation}
 \rm{H^{2}(z)=H_{0}^{2}~(0.27~(1+z)^3+0.73)}
\end{equation}

with gravitational constant $\rm{G\approx4.3\times10^{-3}~pc~M_{\odot}^{-1}~(km/s)^2}$.

\section{Simulated spectra}
\label{sec:simspec}

\begin{figure}
 \centering
 \includegraphics[width=0.44\textwidth,angle=0,trim=0.5cm 0.9cm 0cm 0.9cm,clip=true]{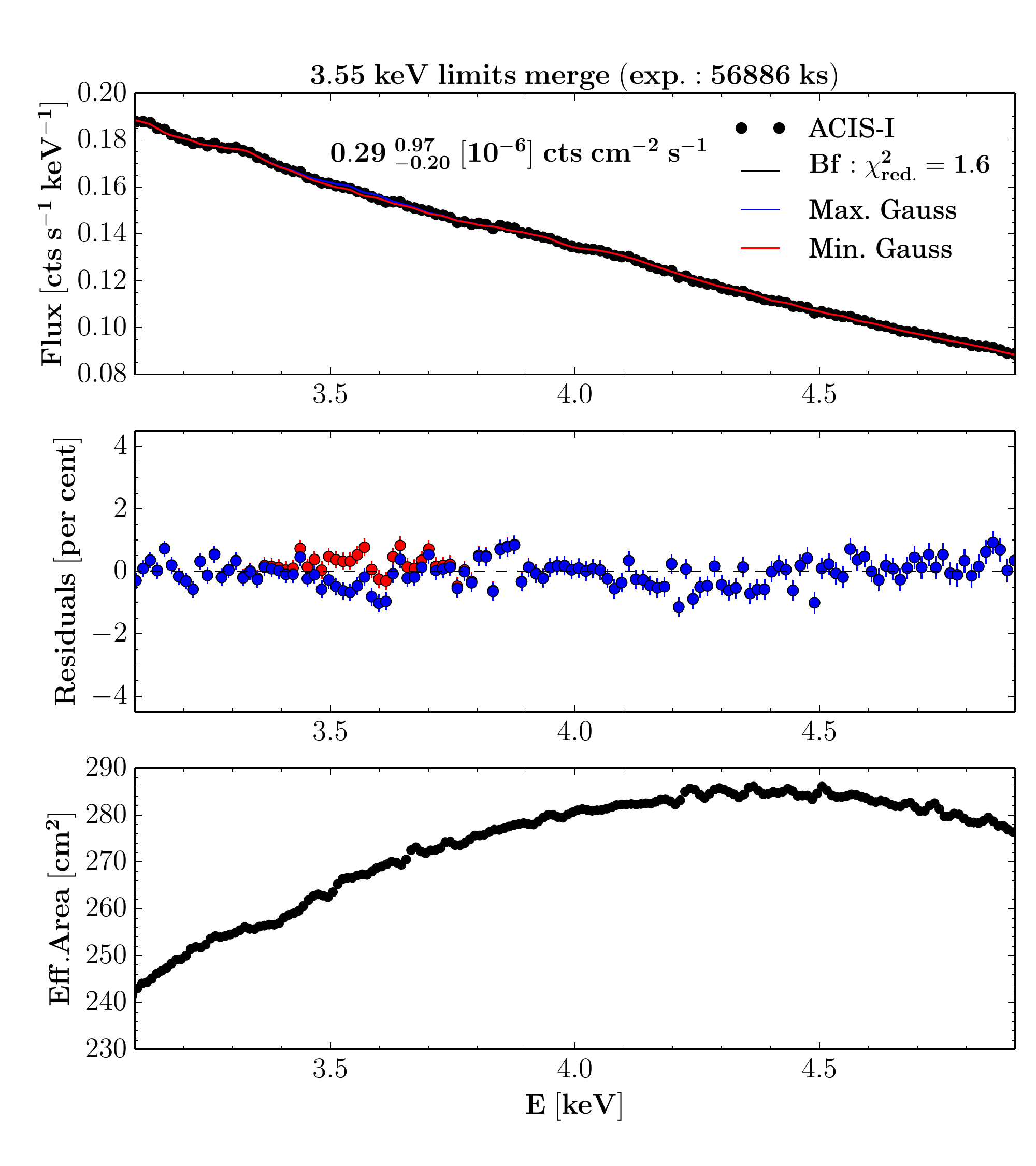}
 \includegraphics[width=0.44\textwidth,angle=0,trim=0.5cm 0.5cm 0cm 0.9cm,clip=true]{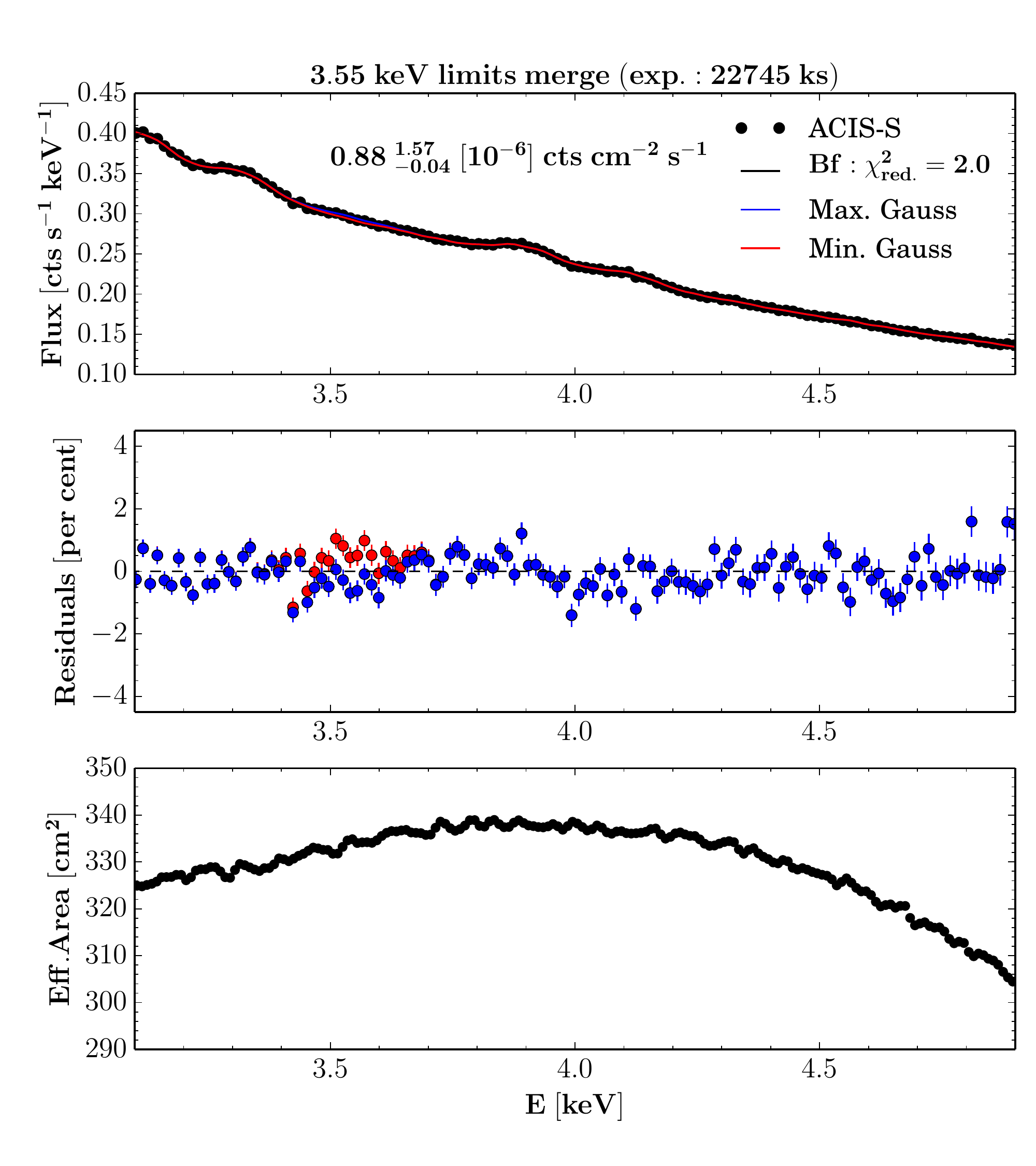}
 \caption{Merged simulations with 10 times the exposure of original X-ray spectra and an additional Gaussian line at 3.55\,keV (ACIS-I top and ACIS-S bottom) with residuals of different fitted models and the effective area (ARF and RMF combined). Fitted \texttt{XSPEC} models: \texttt{apec+apec+zgauss} with best-fit (Bf), upper and lower confidence values (99.7 per cent) of the Gaussian flux in counts $\rm{cm^{-2}~s^{-1}}$. The annotations show the best-fit value and the confidence interval obtained using MCMC. Residuals are shown for the fit with upper (blue) and lower (red) confidence limit of the Gaussian flux.}
 \label{fig:fakelinemerge}
\end{figure}

We used simulated data to test our spectral stacking technique for detecting an additional line at 3.55\,keV. For the simulation we created fake spectra using \texttt{XSPEC}. Simulated spectra were obtained for the best fit model of every individual cluster with ten times the original exposure to get better statistics. We added a Gaussian line with the expected flux for each cluster assuming a mixing angle of $\rm{\sim7 \times 10^{-11}}$ and that all dark matter is made of the 7.1\,keV sterile neutrinos as in previous detections \citep[see][]{2014ApJ...789...13B,2014PhRvL.113y1301B}. Exactly the same merging and modelling procedure was performed on the simulations and the real data. Fig. \ref{fig:fakelinemerge} shows the resulting stacked spectra of the simulation (compare to Fig. \ref{fig:linemerge} for the real data). The additional line is recovered between 95 and 99.7 per cent confidence for both ACIS-I and ACIS-S detectors after the stacking. The measured flux can be converted into a mixing angle using the average sample properties of Tab. \ref{tab:mix}. The resulting mixing angles in the simulations are $\rm{4.1^{+3.2}_{-2.3}}$ (ACIS-I, CI:68) and $\rm{14.8^{+3.9}_{-5.2}}$ (ACIS-S, CI:68).
These simulations show the detection efficiency of our stacking method. There is some hint that the limits of ACIS-S are biased high which is in agreement with the limits obtained from real data (see Fig. \ref{fig:literature} the best constraints). 
There are small fluctuations in the residuals of less than 2 per cent. To asses the uncertainties introduced by our de-redshifting technique of the RMF response files we tested many different combinations of smoothing and number of iterations. The best results were obtained calculating the average of 100 iterations and smoothing them using a Gaussian filter ($\rm{\sigma = 1~pixel}$). This procedure proved most effective in recovering the simulated flux (see Fig. \ref{fig:fakelinemerge}). The residual structures seen in the merged spectra are partially caused by the averaging over many cluster spectra of different temperatures which are not perfectly modelled by the two temperature component \texttt{apec} model. More complex models however made the fit less stable and did not improve the capability to detect the additional line. Another source of residuals can be fluctuations in the average response (see bottom panels Fig. \ref{fig:fakelinemerge} for combined ARF and RMF). These minor fluctuations did not affect our detection efficiency since they averaged out over the width of a spectral line. We extensively tested the detection efficiency using different stacking methods. Additional smoothing of the response files resulted in lower sensitivity for detecting the simulated line (e.g. using mean, median, and Gaussian filters) because it introduced residuals around spectral features due to broadening of the response. Using linear interpolation to remap the response resulted in smoother response but the simulated line was not recovered. The method we used for the final analysis was the most efficient at recovering the simulated line.

\section{Multi-temperature residuals in ACIS-S}
\label{sec:residuals}

When we fit the merged spectrum of the cluster sample for the ACIS-S detectors (Fig. \ref{fig:linemerge}) there were two significant residuals ($\rm{>1\sigma}$). In the simulated merged spectra such strong residuals are not found (Fig. \ref{fig:fakelinemerge}). We simulated each cluster with a two-temperature plasma X-ray emission model (\texttt{XSPEC} \texttt{apec}) and then modelled the merged spectrum with two temperatures (actually the average of $\rm{2\times33}$ temperatures).
The reasons for stronger residuals in the real observations of ACIS-S could be that multi-temperature effects are stronger or that there are additional components not modelled by the current plasma emission model. 
As discussed in the paper there are two significant line-like residuals in ACIS-S at $\rm{\sim3.85~keV}$ and $\rm{\sim4.7~keV}$. The $\rm{\sim3.85~keV}$ residual could be explained by stronger than expected Ca\,XIX emission lines (several lines known between $\rm{\sim3.86-3.90~keV}$). The $\rm{\sim4.7~keV}$ line could be due to Ti\,XXI (see Sect. \ref{sec:discussion}).

\begin{figure}
 \centering
 \includegraphics[width=0.48\textwidth,angle=0,trim=0.7cm 0cm 0cm 0cm,clip=true]{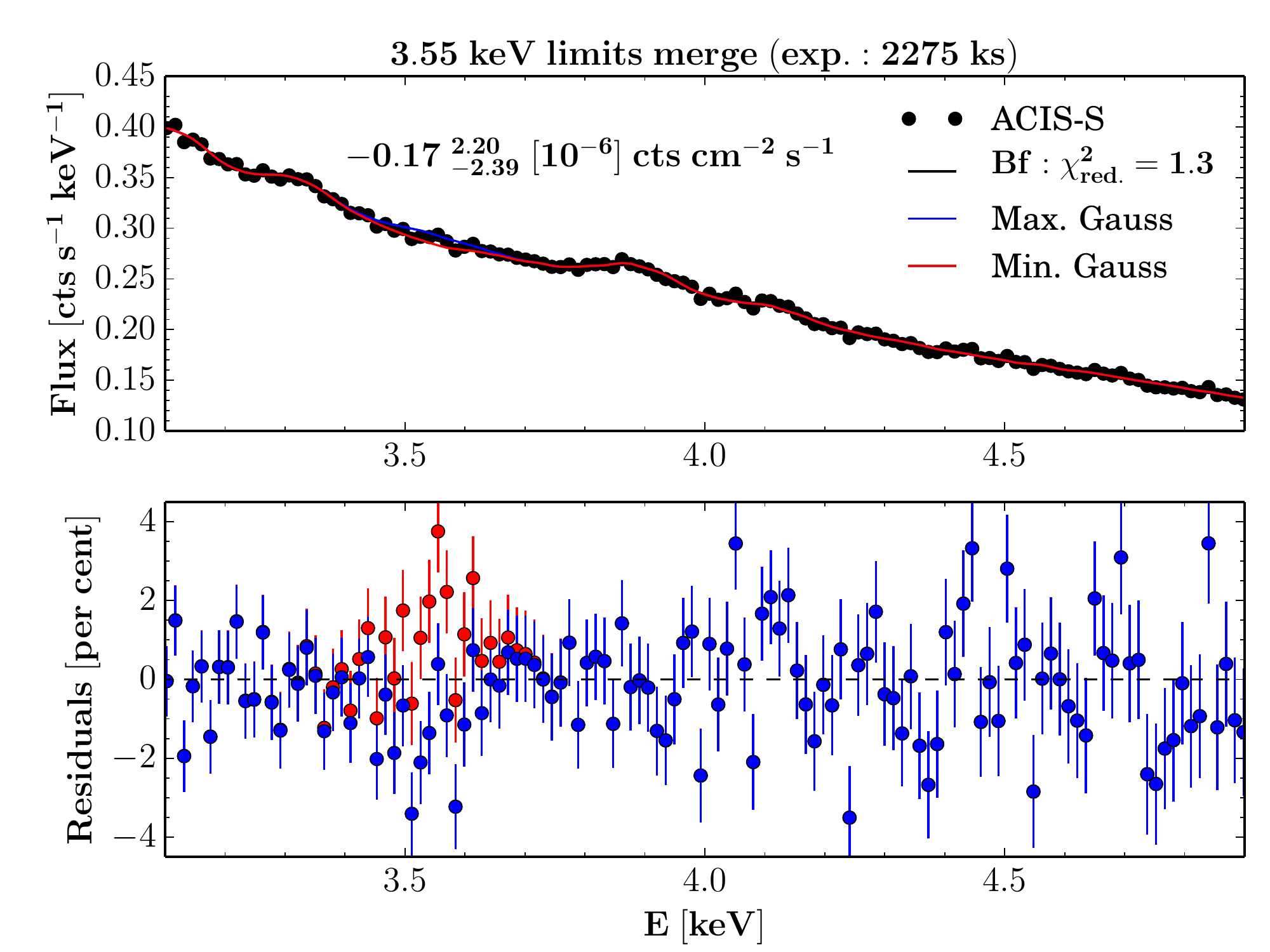}
 \caption{Merged X-ray spectrum (ACIS-S) of the cluster sample with residuals of different fitted models. Fitted \texttt{XSPEC} models: \texttt{apec+apec+zgauss(3.55\,keV) +zgauss(3.85\,keV) +zgauss(4.7\,keV)} with best-fit (Bf), upper and lower confidence values (99.7 per cent) of the Gaussian flux at 3.55\,keV in counts $\rm{cm^{-2}~s^{-1}}$. The annotations show the best-fit value and the confidence interval obtained using MCMC. Residuals are shown for the fit with upper (blue) and lower (red) confidence limit of the 3.55\,keV Gaussian flux. For the effective-area curve see Fig. \ref{fig:fakelinemerge}.}
 \label{fig:acissres}
\end{figure}

To test whether the two temperature fit is sufficient to obtain reliable 3.55\,keV line limits from the merged observations we modelled the two lines with additional Gaussian lines. The line-width was set to zero and the energy set to the two values estimated above. The goodness of fit improved from $\rm{\chi^2_{red.}=1.4}$ to 1.3 and the residual features seen in the original fit disappear (see Fig.\ref{fig:acissres}). Both lines are detected at $\rm{>1\sigma}$.

The obtained limits on the 3.55\,keV flux however are very well consistent with the original fit showing that the additional residuals are negligible for our estimates.

\section{Detailed parameters of fitting procedure}
\label{sec:modeltabs}

\begin{table*}
\caption[]{Best fit \texttt{XSPEC} models.}
\begin{center}
\begin{tabular}{lrrrrrrrr}
\hline\hline\noalign{\smallskip}
  \multicolumn{1}{l}{Cluster\tablefootmark{a}} &
  \multicolumn{1}{l}{T1 [keV]\tablefootmark{b}} &
  \multicolumn{1}{l}{T2 [keV]\tablefootmark{b}} &
  \multicolumn{1}{l}{$\rm{Z~[Z_\odot]}$\tablefootmark{b}} &
  \multicolumn{1}{l}{$\rm{n1~[10^{-2}~cm^{-5}]}$\tablefootmark{b}} &
  \multicolumn{1}{l}{$\rm{n2~[10^{-2}~cm^{-5}]}$\tablefootmark{b}} &
  \multicolumn{1}{l}{$\rm{Chi^{2}_{red}}$\tablefootmark{b}} &
  \multicolumn{1}{l}{DOF\tablefootmark{b}} \\
\noalign{\smallskip}\hline\noalign{\smallskip}

1e0657i & 12.5$\pm$0.48 & - & 0.49$\pm$0.189 & 1.11$\pm$0.031 & - & 1.20 & 200 \\
a1795i & 2.5$\pm$0.35 & 6.4$\pm$0.39 & 0.42$\pm$0.026 & 1.02$\pm$0.348 & 3.57$\pm$0.343 & 1.08 & 199 \\
a1795s & 3.1$\pm$0.45 & 7.4$\pm$0.59 & 0.46$\pm$0.034 & 1.75$\pm$0.454 & 3.01$\pm$0.460 & 1.04 & 199 \\
a1995i & 6.8$\pm$0.97 & - & 0.25$\pm$0.194 & 0.18$\pm$0.120 & - & 1.00 & 84 \\
a1995s & 6.6$\pm$0.88 & - & 0.78$\pm$0.665 & 0.19$\pm$0.129 & - & 1.10 & 87 \\
rxj1347i & 4.2$\pm$1.93 & 16.2$\pm$2.76 & 0.27$\pm$0.050 & 0.23$\pm$0.143 & 1.08$\pm$0.061 & 0.90 & 199 \\
rxj1347s & 11.2$\pm$1.33 & - & 0.31$\pm$0.055 & 1.25$\pm$0.182 & - & 0.90 & 163 \\
zw3146i & 0.5$\pm$0.70 & 6.9$\pm$1.02 & 0.30$\pm$0.295 & 0.90$\pm$4.692 & 1.01$\pm$0.105 & 1.05 & 199 \\
a1835i & 4.1$\pm$2.50 & 8.8$\pm$1.71 & 0.38$\pm$0.207 & 0.36$\pm$0.602 & 1.21$\pm$0.401 & 1.12 & 199 \\
a1835s & 2.1$\pm$1.21 & 11.9$\pm$4.67 & 0.08$\pm$0.151 & 0.75$\pm$0.394 & 1.29$\pm$0.232 & 0.92 & 198 \\
a665i & 4.6$\pm$3.04 & 15.4$\pm$10.09 & 0.39$\pm$0.397 & 0.24$\pm$0.276 & 0.24$\pm$0.306 & 1.03 & 199 \\
a520i & 8.1$\pm$0.30 & - & 0.73$\pm$0.276 & 0.37$\pm$0.027 & - & 1.07 & 200 \\
a1689i & 10.7$\pm$0.80 & - & 0.44$\pm$0.254 & 1.66$\pm$0.075 & - & 1.15 & 200 \\
ms1455i & 3.4$\pm$1.87 & 6.1$\pm$3.24 & 0.10$\pm$0.203 & 0.34$\pm$0.757 & 0.40$\pm$0.767 & 1.07 & 198 \\
a401i & 8.9$\pm$0.42 & - & 0.44$\pm$0.152 & 2.95$\pm$0.089 & - & 1.23 & 200 \\
a1413i & 1.4$\pm$0.77 & 8.9$\pm$1.37 & 0.83$\pm$0.336 & 0.09$\pm$0.137 & 0.96$\pm$0.049 & 1.09 & 199 \\
a2146i & 2.5$\pm$2.73 & 7.1$\pm$1.21 & 0.54$\pm$0.198 & 0.05$\pm$0.184 & 0.48$\pm$0.132 & 0.98 & 199 \\
a2146s & 1.4$\pm$1.03 & 8.6$\pm$2.88 & 0.32$\pm$0.381 & 0.21$\pm$0.296 & 0.47$\pm$0.155 & 0.94 & 169 \\
a521i & 6.4$\pm$1.12 & - & 1.78$\pm$1.961 & 0.08$\pm$0.105 & - & 0.88 & 109 \\
a521s & 8.2$\pm$1.65 & - & 0.01$\pm$1.534 & 0.13$\pm$0.116 & - & 1.15 & 55 \\
ms0735i & 4.9$\pm$0.66 & - & 0.77$\pm$0.239 & 0.34$\pm$0.053 & - & 1.05 & 199 \\
ms0735s & 2.5$\pm$3.72 & 5.5$\pm$2.75 & 0.43$\pm$0.385 & 0.08$\pm$0.524 & 0.31$\pm$0.434 & 1.02 & 157 \\
pks0745i & 2.7$\pm$0.64 & 11.9$\pm$5.97 & 0.68$\pm$0.271 & 1.83$\pm$1.055 & 4.10$\pm$0.616 & 1.03 & 199 \\
pks0745s & 2.7$\pm$0.68 & 9.4$\pm$0.78 & 0.53$\pm$0.074 & 1.27$\pm$0.410 & 5.10$\pm$0.367 & 1.04 & 199 \\
a2204i & 2.0$\pm$0.27 & 13.7$\pm$3.27 & 0.47$\pm$0.157 & 1.03$\pm$0.213 & 2.02$\pm$0.149 & 1.06 & 199 \\
a2204s & 1.7$\pm$0.89 & 8.8$\pm$3.26 & 0.70$\pm$0.511 & 0.56$\pm$0.822 & 2.43$\pm$0.326 & 1.05 & 192 \\
a2034i & 1.4$\pm$0.39 & 12.9$\pm$2.63 & 0.25$\pm$0.122 & 0.21$\pm$0.125 & 0.63$\pm$0.038 & 0.94 & 199 \\
cygnusai & 1.5$\pm$0.42 & 7.4$\pm$0.36 & 0.86$\pm$0.111 & 0.31$\pm$0.158 & 3.64$\pm$0.094 & 0.90 & 199 \\
cygnusas & 2.5$\pm$0.40 & 11.6$\pm$3.47 & 0.68$\pm$0.145 & 1.32$\pm$0.450 & 2.98$\pm$0.265 & 0.99 & 199 \\
a907i & 3.7$\pm$2.94 & 7.2$\pm$2.90 & 0.40$\pm$0.218 & 0.33$\pm$0.756 & 0.41$\pm$0.710 & 0.88 & 199 \\
a3667i & 3.2$\pm$0.50 & 10.6$\pm$1.37 & 0.22$\pm$0.065 & 0.92$\pm$0.226 & 1.54$\pm$0.207 & 0.90 & 199 \\
2a0335s & 1.4$\pm$0.39 & 3.4$\pm$0.24 & 0.67$\pm$0.042 & 1.69$\pm$0.831 & 5.23$\pm$0.841 & 1.41 & 199 \\
a2597s & 0.8$\pm$1.28 & 3.7$\pm$0.10 & 0.43$\pm$0.058 & 0.14$\pm$0.668 & 1.91$\pm$0.063 & 0.98 & 199 \\
a1650i & 2.8$\pm$2.31 & 7.3$\pm$1.10 & 0.45$\pm$0.102 & 0.32$\pm$0.560 & 1.25$\pm$0.377 & 0.85 & 199 \\
a1650s & 4.3$\pm$1.63 & 6.9$\pm$2.96 & 0.63$\pm$0.205 & 0.81$\pm$1.128 & 0.78$\pm$1.141 & 1.09 & 199 \\
a2199i & 3.0$\pm$0.49 & 5.9$\pm$1.02 & 0.41$\pm$0.048 & 3.35$\pm$1.251 & 2.91$\pm$1.271 & 0.89 & 199 \\
a2199s & 2.0$\pm$0.36 & 6.1$\pm$1.01 & 0.38$\pm$0.058 & 2.23$\pm$1.003 & 4.74$\pm$0.895 & 0.96 & 199 \\
hydraai & 3.7$\pm$0.14 & - & 0.83$\pm$0.599 & 2.54$\pm$0.428 & - & 1.55 & 199 \\
hydraas & 1.4$\pm$1.00 & 4.0$\pm$0.22 & 0.38$\pm$0.036 & 0.36$\pm$0.467 & 3.42$\pm$0.333 & 1.06 & 199 \\
a496s & 2.1$\pm$0.14 & 7.1$\pm$0.59 & 0.87$\pm$0.082 & 1.10$\pm$0.223 & 3.32$\pm$0.172 & 1.06 & 199 \\
sersic159i & 2.2$\pm$0.27 & 3.9$\pm$1.94 & 0.38$\pm$0.058 & 1.28$\pm$0.752 & 0.58$\pm$0.916 & 1.14 & 199 \\
sersic159s & 2.0$\pm$0.81 & 4.3$\pm$5.23 & 0.46$\pm$0.154 & 1.17$\pm$1.722 & 0.66$\pm$1.958 & 1.12 & 131 \\
3c348s & 2.6$\pm$1.11 & 4.7$\pm$1.62 & 0.75$\pm$0.221 & 0.12$\pm$0.385 & 0.27$\pm$0.322 & 0.98 & 196 \\
a1775s & 2.3$\pm$1.12 & 5.7$\pm$3.89 & 0.57$\pm$0.149 & 0.20$\pm$0.346 & 0.19$\pm$0.339 & 1.08 & 191 \\
a2052s & 1.5$\pm$0.23 & 3.4$\pm$0.12 & 0.67$\pm$0.029 & 0.48$\pm$0.171 & 1.96$\pm$0.126 & 1.02 & 199 \\
a2744i & 10.2$\pm$1.65 & - & 0.31$\pm$0.359 & 0.46$\pm$0.093 & - & 1.02 & 200 \\
a2744s & 10.9$\pm$2.52 & - & 0.93$\pm$1.381 & 0.43$\pm$0.268 & - & 0.87 & 122 \\
a2390s & 5.4$\pm$5.40 & 10.3$\pm$2.40 & 0.95$\pm$0.305 & 0.26$\pm$0.998 & 1.02$\pm$0.612 & 0.92 & 199 \\
mergei & 2.8$\pm$0.85 & 7.3$\pm$0.34 & 0.56$\pm$0.099 & 0.30$\pm$0.165 & 1.54$\pm$0.130 & 1.11 & 130 \\
merges & 2.2$\pm$0.21 & 5.8$\pm$0.32 & 0.70$\pm$0.063 & 0.97$\pm$0.182 & 2.17$\pm$0.155 & 1.41 & 130 \\

\noalign{\smallskip}\hline
\end{tabular}
\tablefoot{
\tablefoottext{a}{Abbreviated cluster name and detector identifier (i: ACIS-I, s: ACIS-s). ``merge`` indicates the best fit to the merged spectra of the complete sample.}
\tablefoottext{b}{Best fit parameters of the \texttt{XSPEC} \texttt{apec} models to the individual cluster emission. Second temperature model was set to zero if temperatures T differed by less than 30 per cent or second normalisation n was less than one per cent of the first model. Metallicity Z was linked between the two models. The goodness of fit is give as $\rm{Chi^{2}_{red}}$ with the degrees of freedom (DOF) of the fit.}
}
\end{center}
\label{tab:xspecpars}
\end{table*}

\begin{table*}
\caption[]{Fixed cluster parameters and weights.}
\begin{center}
\begin{tabular}{lrrrrrrrrr}
\hline\hline\noalign{\smallskip}
  \multicolumn{1}{l}{Cluster\tablefootmark{a}} &
  \multicolumn{1}{l}{z} &
  \multicolumn{1}{l}{$\rm{n_H}$} &
  \multicolumn{1}{l}{Expw.i} &
  \multicolumn{1}{l}{Expw.s} &
  \multicolumn{1}{l}{$\rm{\omega}$.i} &
  \multicolumn{1}{l}{$\rm{\omega}$.s} &
  \multicolumn{1}{l}{$\rm{M_{DM}^{FOV}}$} &
  \multicolumn{1}{l}{$\rm{{M_{DM}^{FOV}}/D^2}$} \\[1pt]
  \multicolumn{1}{l}{} &
  \multicolumn{1}{l}{} &
  \multicolumn{1}{l}{$\rm{[10^{22}~cm^{-2}]}$} &
  \multicolumn{1}{l}{[per cent]\tablefootmark{b}} &
  \multicolumn{1}{l}{[per cent]} &
  \multicolumn{1}{l}{} &
  \multicolumn{1}{l}{} &
  \multicolumn{1}{l}{$\rm{[10^{14}~M_\odot]}$\tablefootmark{c}} &
  \multicolumn{1}{l}{$\rm{[10^{10}~M_\odot~Mpc^{-2}]}$} \\
\noalign{\smallskip}\hline\noalign{\smallskip}
1e0657 & 0.30 &  0.049 &  10.0 &   - &   3.5 &   - &   5.92 & 0.026 \\
a1795 & 0.06 &  0.012 &  10.8 &   10.6 &   25.5 &   19.3 &   1.59 & 0.210 \\
a1995 & 0.32 &  0.012 &  1.0 &   2.0 &   0.1 &   0.1 &   1.66 & 0.006 \\
rxj1347 & 0.45 &  0.046 &  3.8 &   0.8 &   0.6 &   0.1 &   6.50 & 0.010 \\
zw3146 & 0.28 &  0.025 &  1.5 &   - &   0.2 &   - &   2.49 & 0.012 \\
a1835 & 0.25 &  0.020 &  3.4 &   1.3 &   1.0 &   0.3 &   3.38 & 0.021 \\
a665 & 0.18 &  0.043 &  2.4 &   - &   0.8 &   - &   2.03 & 0.027 \\
a520 & 0.20 &  0.057 &  9.4 &   - &   1.3 &   - &   1.07 & 0.011 \\
a1689 & 0.18 &  0.018 &  3.5 &   - &   2.2 &   - &   3.99 & 0.051 \\
ms1455 & 0.26 &  0.032 &  1.9 &   - &   0.2 &   - &   1.26 & 0.008 \\
a401 & 0.07 &  0.099 &  2.9 &   - &   8.0 &   - &   2.69 & 0.247 \\
a1413 & 0.14 &  0.018 &  2.4 &   - &   1.6 &   - &   2.39 & 0.054 \\
a2146 & 0.23 &  0.030 &  6.7 &   1.9 &   1.0 &   0.2 &   1.51 & 0.011 \\
a521 & 0.25 &  0.049 &  2.3 &   1.7 &   0.2 &   0.1 &   1.15 & 0.008 \\
ms0735 & 0.21 &  0.033 &  8.5 &   2.0 &   1.5 &   0.3 &   1.51 & 0.014 \\
pks0745 & 0.10 &  0.373 &  0.3 &   7.0 &   0.4 &   8.7 &   3.03 & 0.138 \\
a2204 & 0.15 &  0.057 &  1.5 &   0.4 &   1.1 &   0.3 &   3.10 & 0.061 \\
a2034 & 0.11 &  0.015 &  4.6 &   - &   4.3 &   - &   2.14 & 0.080 \\
cygnusa & 0.06 &  0.272 &  3.5 &   1.5 &   8.5 &   2.9 &   1.33 & 0.218 \\
a907 & 0.17 &  0.054 &  1.8 &   - &   0.5 &   - &   1.44 & 0.023 \\
a3667 & 0.06 &  0.044 &  8.3 &   - &   22.3 &   - &   1.44 & 0.240 \\
2a0335 & 0.03 &  0.175 &  - &   4.5 &   - &   7.1 &   0.42 & 0.187 \\
a2597 & 0.09 &  0.025 &  - &   6.2 &   - &   2.6 &   0.70 & 0.047 \\
a1650 & 0.08 &  0.013 &  3.6 &   1.2 &   3.8 &   1.0 &   1.35 & 0.093 \\
a2199 & 0.03 &  0.009 &  2.1 &   1.6 &   9.3 &   5.4 &   0.68 & 0.406 \\
hydraa & 0.05 &  0.047 &  0.4 &   8.8 &   0.6 &   9.8 &   0.73 & 0.130 \\
a496 & 0.03 &  0.038 &  - &   3.9 &   - &   9.8 &   0.60 & 0.299 \\
sersic159 & 0.06 &  0.011 &  1.7 &   0.4 &   1.1 &   0.2 &   0.35 & 0.057 \\
3c348 & 0.15 &  0.062 &  - &   5.0 &   - &   0.7 &   0.77 & 0.015 \\
a1775 & 0.07 &  0.010 &  - &   4.4 &   - &   1.4 &   0.38 & 0.037 \\
a2052 & 0.04 &  0.027 &  - &   28.7 &   - &   27.9 &   0.27 & 0.115 \\
a2744 & 0.31 &  0.014 &  1.8 &   1.1 &   0.3 &   0.1 &   2.57 & 0.010 \\
a2390 & 0.23 &  0.062 &  - &   4.9 &   - &   1.9 &   5.14 & 0.039 \\

\noalign{\smallskip}\hline
\end{tabular}
\tablefoot{
\tablefoottext{a}{Abbreviated cluster name, redshift z, and Galactic foreground absorption from previous catalogs \citep[for convention see][]{2016A&A...585A.130H}.}
\tablefoottext{b}{Expw: fraction of total exposure in per cent; $\rm{\omega}$: weighting factor for expected contribution to the DM line flux (i: ACIS-I, s: ACIS-S) see equation (\ref{eqn:weights}).}
\tablefoottext{c}{Averaged DM mass in the FOV, and the same mass divided by the luminosity distance of the cluster (main indicator of expected DM line flux).}
}
\end{center}
\label{tab:clusterpars}
\end{table*}

\begin{table*}
\caption[]{Flux and mixing angle constraints.}
\begin{center}
\begin{tabular}{lrrrrrrr}
\hline\hline\noalign{\smallskip}
  \multicolumn{1}{l}{Cluster\tablefootmark{a}} &
  \multicolumn{1}{l}{$\rm{F_X}$} &
  \multicolumn{1}{l}{$\rm{F_X^{up}}$} &
  \multicolumn{1}{l}{$\rm{F_X^{lo}}$} &
  \multicolumn{1}{l}{$\rm{sin^2(2\Theta)}$} &
  \multicolumn{1}{l}{$\rm{sin^2(2\Theta)^{up}}$} &
  \multicolumn{1}{l}{$\rm{sin^2(2\Theta)^{lo}}$} \\
  \multicolumn{1}{l}{} &
  \multicolumn{1}{l}{$\rm{[10^{-6}~cts~cm^{-2}~s^{-1}]}$\tablefootmark{b}} &
  \multicolumn{1}{l}{} &
  \multicolumn{1}{l}{} &
  \multicolumn{1}{l}{$\rm{[10^{-9}]}$\tablefootmark{c}} &
  \multicolumn{1}{l}{} &
  \multicolumn{1}{l}{} \\
\noalign{\smallskip}\hline\noalign{\smallskip}

1e0657i & 0.13 & 3.75 & -4.08  & 0.12 & 3.48 & -3.79 \\
a1795i & -1.03 & 3.22 & -5.66  & -0.14 & 0.44 & -0.78 \\
a1795s & -1.04 & 4.73 & -6.46  & -0.14 & 0.65 & -0.89 \\
a1995i & 0.39 & 5.43 & -4.48  & 1.51 & 20.77 & -17.15 \\
a1995s & 0.42 & 5.27 & -4.66  & 1.59 & 20.14 & -17.84 \\
rxj1347i & 1.67 & 7.27 & -4.35  & 3.41 & 14.83 & -8.87 \\
rxj1347s & -7.79 & 9.05 & -24.55  & -15.88 & 18.45 & -50.07 \\
zw3146i & 2.12 & 10.11 & -8.25  & 4.31 & 20.59 & -16.79 \\
a1835i & -2.23 & 6.42 & -8.17  & -2.60 & 7.48 & -9.52 \\
a1835s & 4.05 & 13.66 & -11.45  & 4.71 & 15.92 & -13.34 \\
a665i & 2.40 & 6.93 & -0.71  & 2.35 & 6.78 & -0.70 \\
a520i & 0.01 & 2.23 & -2.05  & 0.02 & 5.23 & -4.81 \\
a1689i & 5.56 & 12.59 & -0.22  & 2.81 & 6.37 & -0.11 \\
ms1455i & -1.54 & 2.58 & -6.92  & -5.04 & 8.42 & -22.61 \\
a401i & -2.09 & 7.71 & -12.51  & -0.24 & 0.90 & -1.46 \\
a1413i & 1.51 & 7.88 & -4.42  & 0.76 & 3.97 & -2.23 \\
a2146i & -0.38 & 2.24 & -3.11  & -0.85 & 4.95 & -6.89 \\
a2146s & -2.91 & 3.20 & -10.82  & -6.44 & 7.10 & -23.97 \\
a521i & -0.22 & 1.88 & -2.48  & -0.73 & 6.19 & -8.15 \\
a521s & -1.51 & 2.22 & -4.16  & -4.98 & 7.30 & -13.68 \\
ms0735i & 0.98 & 1.12 & -0.45  & 1.82 & 2.09 & -0.84 \\
ms0735s & -3.44 & -0.05 & -6.04  & -6.41 & -0.09 & -11.23 \\
pks0745i & 1.15 & 34.50 & -34.77  & 0.23 & 6.99 & -7.05 \\
pks0745s & -2.68 & 7.50 & -13.27  & -0.54 & 1.52 & -2.69 \\
a2204i & 3.49 & 14.83 & -9.40  & 1.53 & 6.51 & -4.13 \\
a2204s & 3.45 & 29.91 & -22.58  & 1.51 & 13.13 & -9.91 \\
a2034i & -0.13 & 2.76 & -3.29  & -0.05 & 0.96 & -1.14 \\
cygnusai & -8.00 & -0.21 & -16.11  & -1.07 & -0.03 & -2.16 \\
cygnusas & -2.69 & 12.85 & -19.01  & -0.36 & 1.72 & -2.55 \\
a907i & 1.74 & 4.55 & -4.02  & 2.01 & 5.26 & -4.64 \\
a3667i & -0.80 & 2.54 & -6.72  & -0.10 & 0.31 & -0.82 \\
2a0335s & 3.11 & 12.07 & -7.77  & 0.50 & 1.92 & -1.24 \\
a2597s & -0.32 & 4.53 & -5.99  & -0.19 & 2.72 & -3.60 \\
a1650i & -0.05 & 5.82 & -5.45  & -0.02 & 1.77 & -1.66 \\
a1650s & -5.96 & 3.08 & -15.21  & -1.81 & 0.94 & -4.63 \\
a2199i & -0.85 & 9.69 & -17.98  & -0.06 & 0.72 & -1.33 \\
a2199s & -0.46 & 12.89 & -18.15  & -0.03 & 0.95 & -1.34 \\
hydraai & 28.58 & 64.89 & -10.65  & 6.45 & 14.64 & -2.40 \\
hydraas & -0.02 & 4.96 & -4.74  & -0.00 & 1.12 & -1.07 \\
a496s & -3.11 & 7.55 & -13.04  & -0.31 & 0.76 & -1.30 \\
sersic159i & -0.37 & 4.77 & -6.64  & -0.19 & 2.43 & -3.39 \\
sersic159s & 9.45 & 24.14 & -6.22  & 4.82 & 12.31 & -3.17 \\
3c348s & 1.08 & 3.29 & -1.90  & 1.99 & 6.04 & -3.49 \\
a1775s & 0.98 & 3.60 & -1.26  & 0.77 & 2.83 & -0.99 \\
a2052s & -1.03 & 1.36 & -3.57  & -0.27 & 0.35 & -0.92 \\
a2744i & -1.70 & 2.46 & -6.98  & -3.90 & 5.65 & -16.04 \\
a2744s & 0.16 & 9.70 & -10.34  & 0.36 & 22.28 & -23.74 \\
a2390s & 2.00 & 8.71 & -4.40  & 1.29 & 5.62 & -2.84 \\

\noalign{\smallskip}\hline
\end{tabular}
\tablefoot{
\tablefoottext{a}{Abbreviated cluster name and detector identifier (i: ACIS-I, s: ACIS-s).}
\tablefoottext{b}{Best fit and $\rm{3\sigma}$ upper and lower boundaries of a possible additional flux added by a Gaussian line at 3.55\,keV.}
\tablefoottext{c}{Mixing-angle limits ($\rm{3\sigma}$ confidence) for the 7.1\,keV sterile neutrino decay scenario (negative mixing-angle unphysical, results from allowing negative flux).}
}
\end{center}
\label{tab:indi_limits}
\end{table*}

\section{3.55\,keV spectra}
\label{sec:alim}

\begin{figure*}
  \resizebox{\hsize}{!}{\includegraphics[page=1,angle=0,trim=1cm 3cm 0cm 1cm,clip=true]{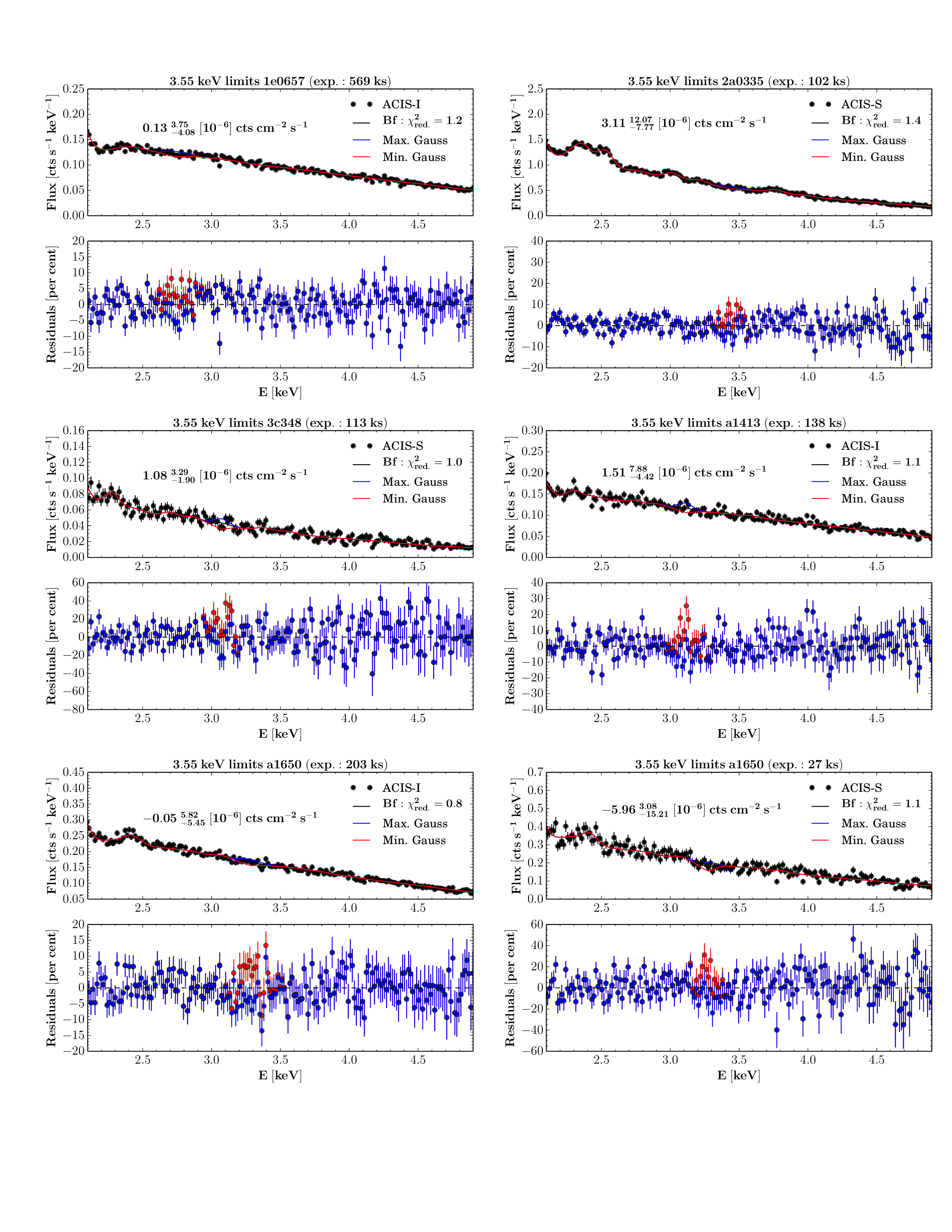}}
  \caption{X-ray spectra (ACIS-I and ACIS-S) with residuals of different fitted models. Fitted models: \texttt{XSPEC} \texttt{phabs*(apec+apec+zgauss)} with best-fit, upper and lower confidence values (99.7 per cent) of the Gaussian flux in $\rm{cm^{-2}~s^{-1}}$. The annotations show the best-fit (Bf) value and the confidence interval obtained using Monte Carlo Markov Chains (MCMC). Residuals are shown for the fit with upper (blue) and lower (red) confidence limit of the Gaussian flux. The plot title indicates the abbreviated cluster name and the exposure time of the observation.}
  \label{fig:3.55all}
\end{figure*}
\addtocounter{figure}{-1} 
\begin{figure*}
  \resizebox{\hsize}{!}{\includegraphics[page=1,angle=0,trim=1cm 3cm 0cm 1cm,clip=true]{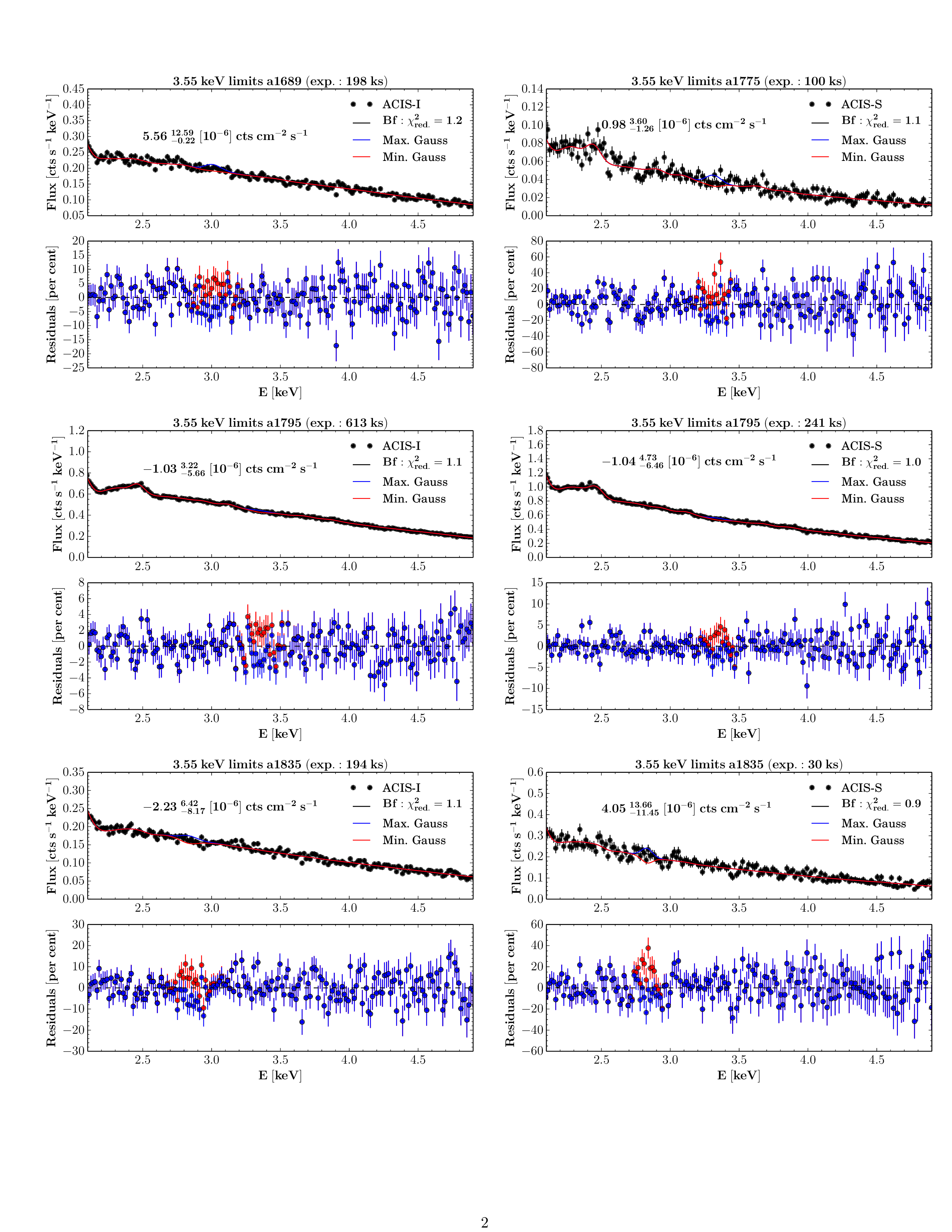}}
  \caption{continued.}
\end{figure*}
\addtocounter{figure}{-1} 
\begin{figure*}
  \resizebox{\hsize}{!}{\includegraphics[page=1,angle=0,trim=1cm 3cm 0cm 1cm,clip=true]{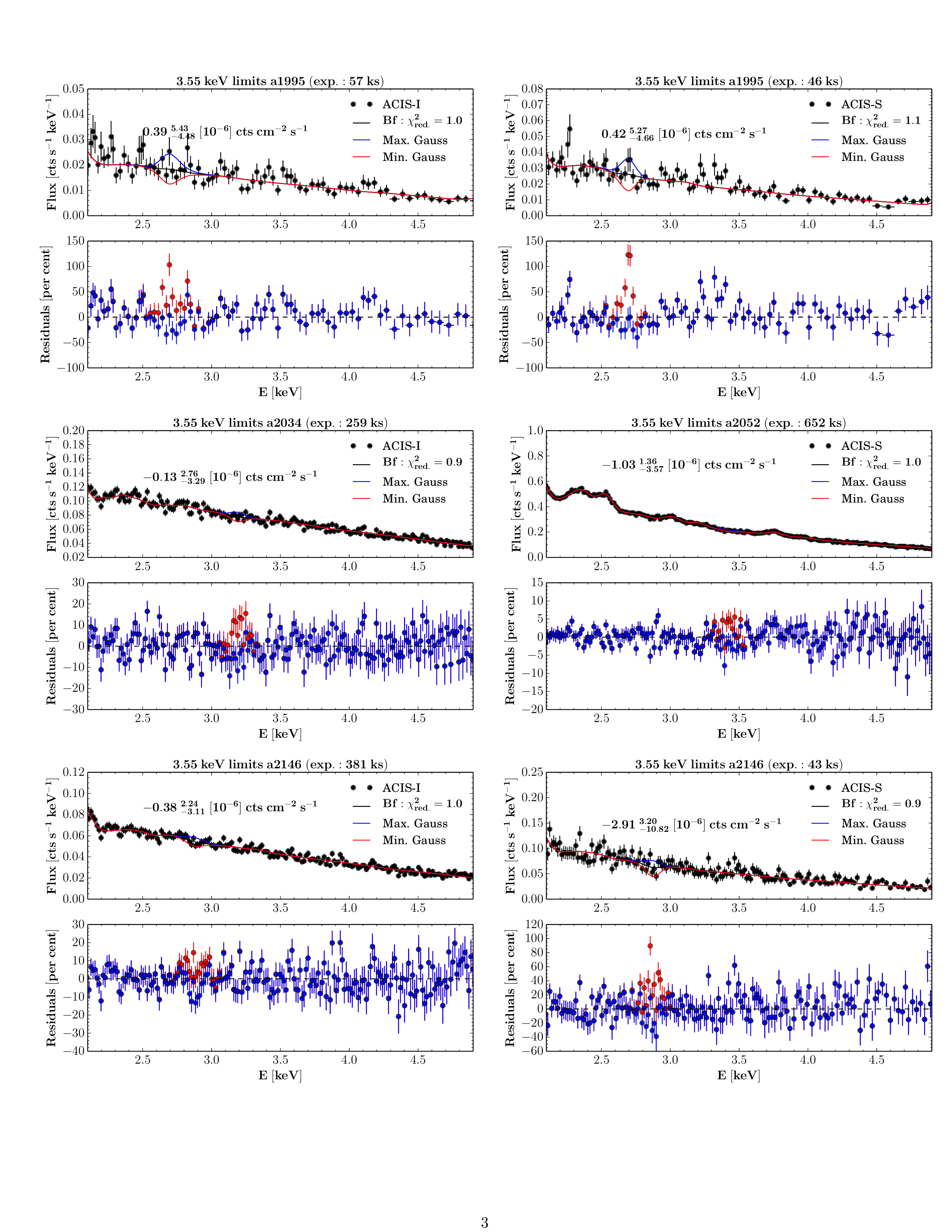}}
  \caption{continued.}
\end{figure*}
\addtocounter{figure}{-1} 
\begin{figure*}
  \resizebox{\hsize}{!}{\includegraphics[page=1,angle=0,trim=1cm 3cm 0cm 1cm,clip=true]{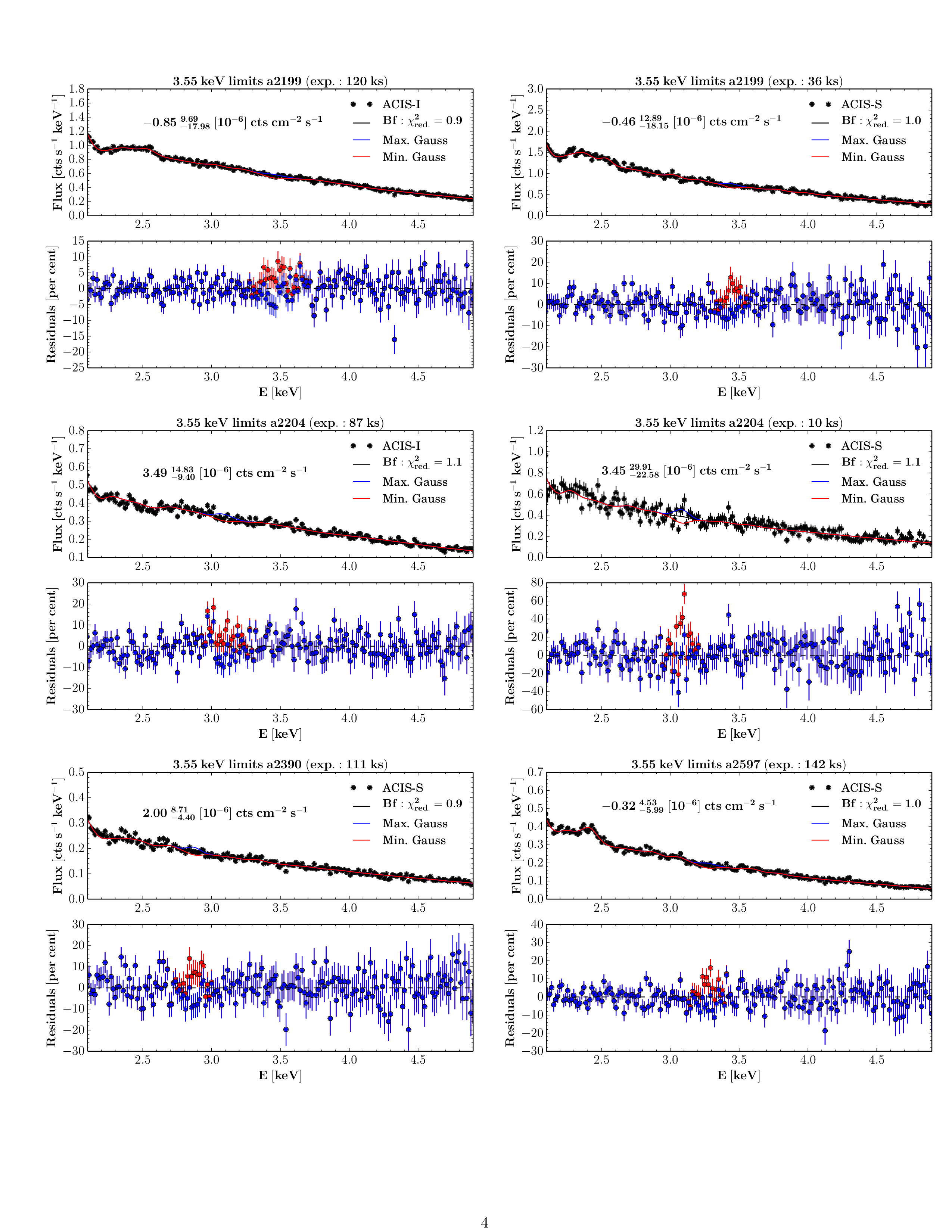}}
  \caption{continued.}
\end{figure*}
\addtocounter{figure}{-1} 
\begin{figure*}
  \resizebox{\hsize}{!}{\includegraphics[page=1,angle=0,trim=1cm 3cm 0cm 1cm,clip=true]{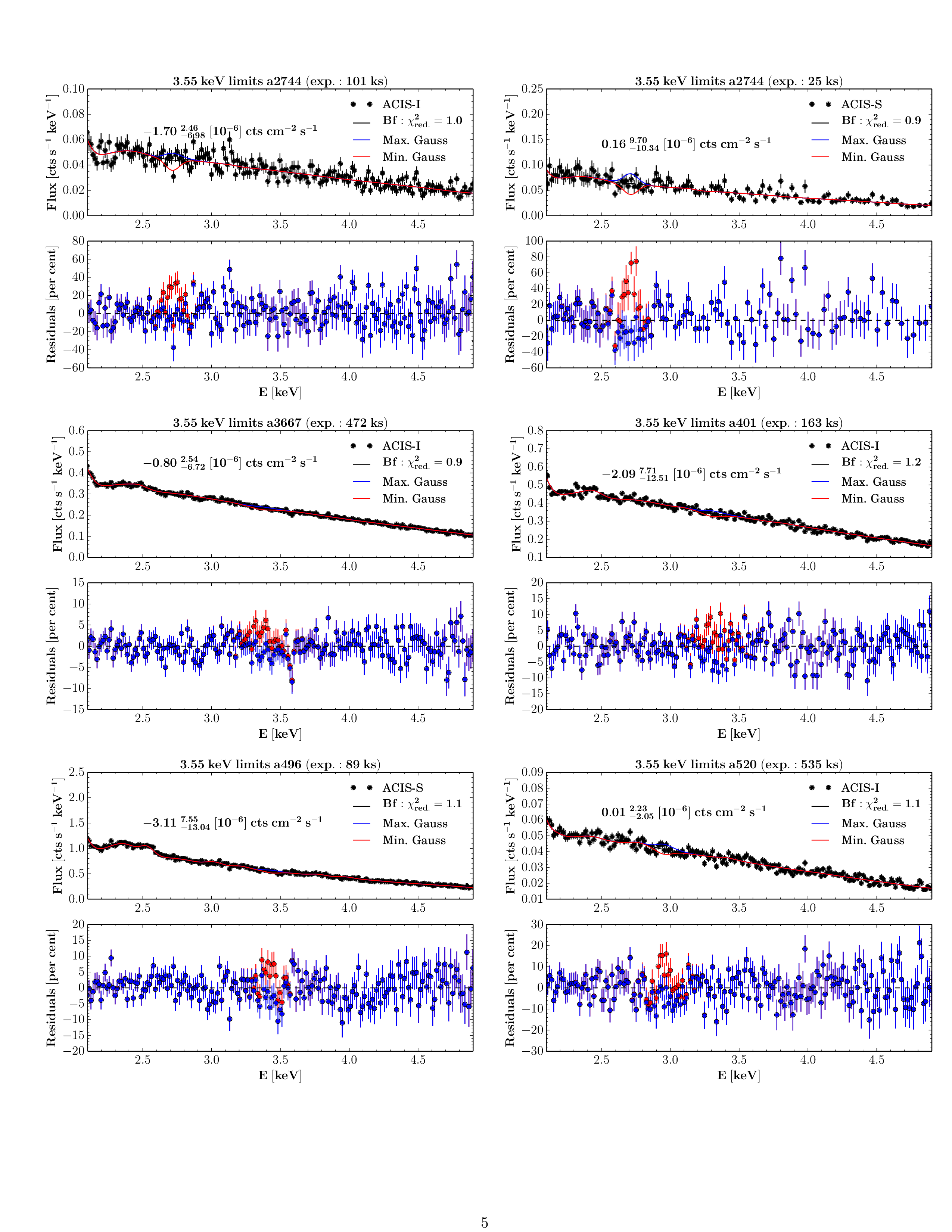}}
  \caption{continued.}
\end{figure*}
\addtocounter{figure}{-1} 
\begin{figure*}
  \resizebox{\hsize}{!}{\includegraphics[page=1,angle=0,trim=1cm 3cm 0cm 1cm,clip=true]{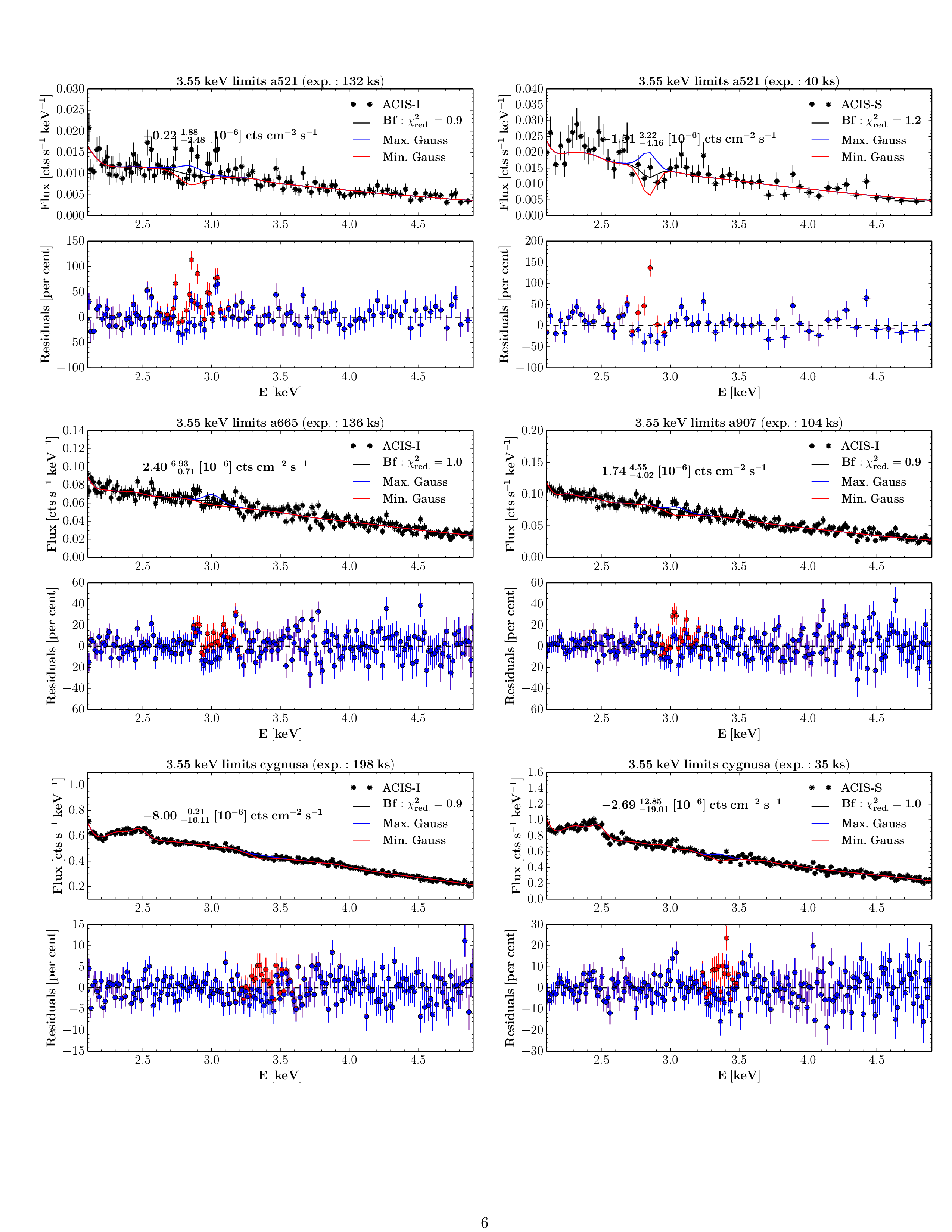}}
  \caption{continued.}
\end{figure*}
\addtocounter{figure}{-1} 
\begin{figure*}
  \resizebox{\hsize}{!}{\includegraphics[page=1,angle=0,trim=1cm 3cm 0cm 1cm,clip=true]{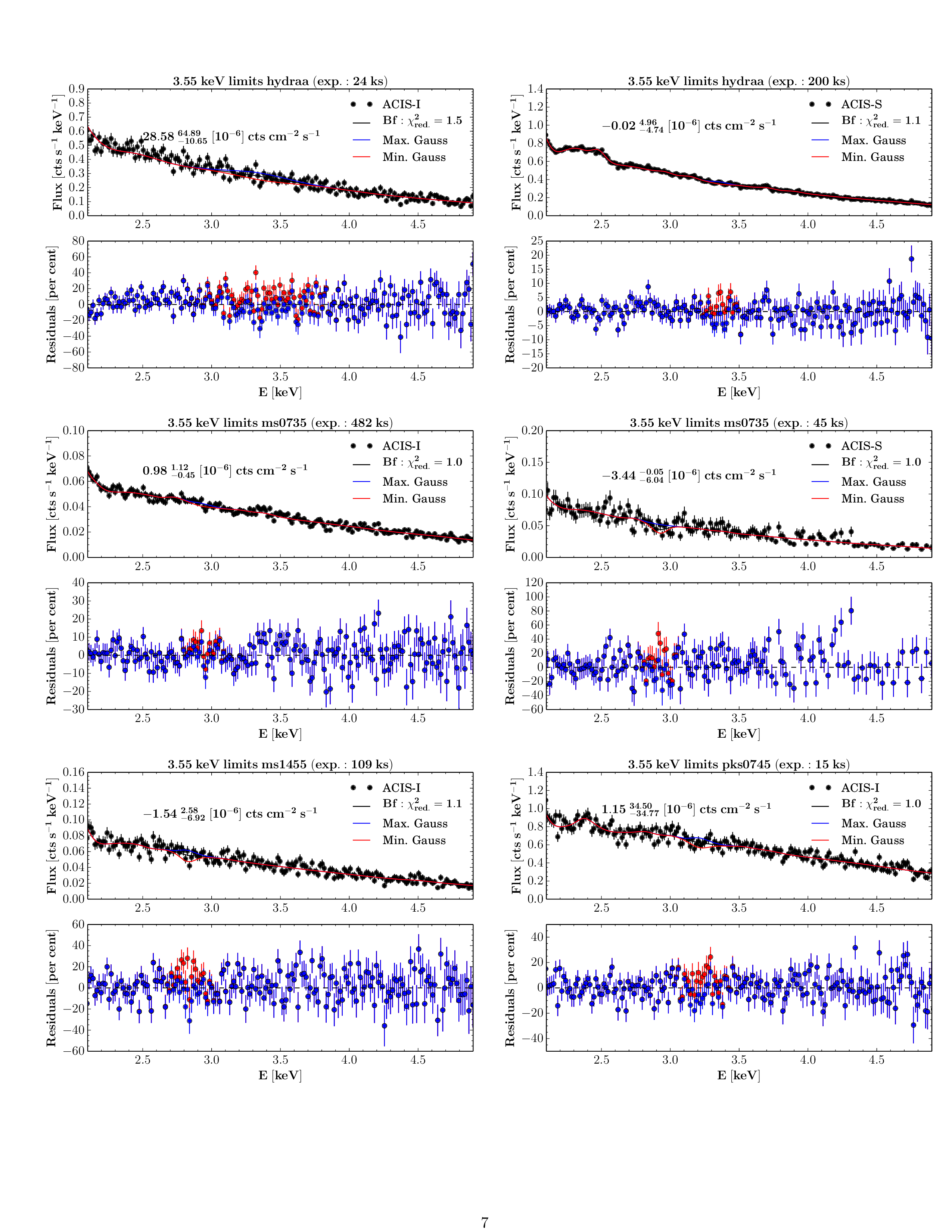}}
  \caption{continued.}
\end{figure*}
\addtocounter{figure}{-1} 
\begin{figure*}
  \resizebox{\hsize}{!}{\includegraphics[page=1,angle=0,trim=0.5cm 3cm 0cm 1cm,clip=true]{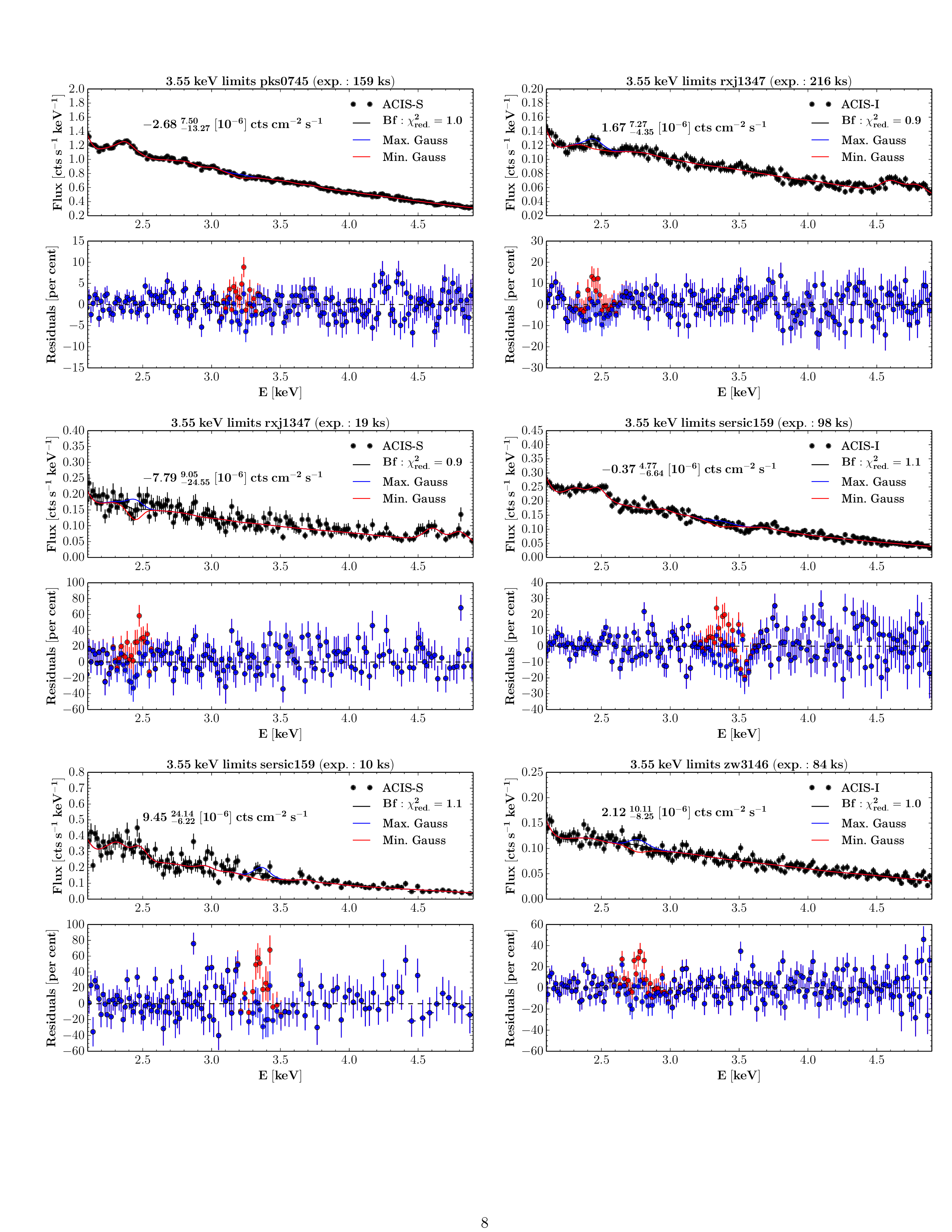}}
  \caption{continued.}
\end{figure*}

\end{appendix}

\end{document}